\documentstyle[epsbox,seceq,onecolumn]{jpsj_sissa}
\newcommand{\eps}{\varepsilon}
\newcommand{\vphi}{\varphi}
\newcommand{\dis}{\displaystyle}
\newcommand{\Mbar}{\bar{M}}
\newcommand{\Dbar}{\bar{\Delta}}
\newcommand{\Deff}{\Delta_{\rm eff}}
\newcommand{\gbar}{\bar{\gamma}}
\def\runtitle{Thermodynamics and Optical Conductivity
of a Dissipative Carrier}
\def\runauthor{Takeo {\sc Kato} and Masatoshi {\sc Imada}}
\title
{Thermodynamics and Optical Conductivity
of a Dissipative Carrier \\ in a Tight Binding Model}

\author
{Takeo Kato and Masatoshi Imada} 

\inst
{Institute for Solid State Physics, 
University of Tokyo,\\ 7-22-1 Roppongi, Minato-ku, Tokyo 106}

\recdate
{November 5,1997}

\abst{
Thermodynamics and transport properties 
of a dissipative particle in a tight-binding model
are studied through specific heat and optical conductivity.
A weak coupling theory is constituted to
study the crossover behavior between the low-temperature
region and the high-temperature region analytically.
We found that coherent part around zero frequency
in the optical conductivity disappears for $0<s<2$, 
where $s$ is an exponent of a spectral
function of the environment.
Detailed calculation is performed for ohmic damping ($s=1$).
In this case, the specific heat shows an unusual 
$T$-linear behavior at low temperatures, which
indicates that the environment strongly influences
the particle motion, and changes the low-energy states 
of the dissipative particle. 
The optical conductivity $\sigma(\omega)$ 
takes a non-Drude form even at zero temperature,
and the high-frequency side behaves as $\omega^{2K-2}$,
where $K$ is a dimensionless damping strength.
The high frequency side of the optical conductivity
is independent of temperatures, while the low frequency side
depends on the temperature, and behaves as $T^{2K-2}$ 
at high temperatures. We also comment on the
application of this model to the description of incoherent
motion in correlated electron systems.
}

\kword
{dissipation, specific heat, optical conductivity, 
ohmic damping, Kubo formula, coherence, incoherence}
\begin{document}
\sloppy
\maketitle
%
%

\section{Introduction}

The dissipative dynamics of a quantum 
particle coupled to a heat bath
has been studied for several decades in many areas
of physics.~\cite{Weiss93} The dissipative dynamics has been
studied theoretically on a phenomenological model
in which a particle is coupled to harmonic oscillators.
This phenomenological model is called
the Caldeira-Leggett model. The model has been used
to study tunneling effects of dissipative phase motion
in Josephson junctions.~\cite{Caldeira83} 
In the Caldeira-Leggett model,
a dissipative particle in a potential $V(q)$ is described
by the Hamiltonian
\begin{eqnarray}
   H &=& \frac{p^2}{2M} + V(q) + \sum_j \left[
   \frac{p_j^2}{2m_j} + \frac12 m_j \omega_j^2 \left(
   x_j - \frac{C_j}{m_j\omega_j^2} q \right)^2 \right],
   \label{Ham} \\
\end{eqnarray}
where $\{q_j\}$ and $\{p_j\}$ are sets of positions and momenta
of harmonic oscillators. The influence of the environment
is determined through the spectral density
\begin{equation}
   J(\omega) = \frac{\pi}{2} \sum_j \frac{C_j^2}{m_j \omega_j}
   \delta(\omega-\omega_j).
\end{equation}
By using the real-time path integral method,
the classical equation of motion of a dissipative particle 
is derived as~\cite{Weiss93,Schmid82}
\begin{equation}
M \ddot{q}(t) + M \int_{-\infty}^{t} {\rm d}t' \, 
\gamma (t-t') \dot q (t') + \frac{\partial V}{\partial q}
= \xi(t).
\end{equation}
Here, $\xi(t)$ is a random force, and $\gamma(t)$
is a damping kernel determined by
\begin{equation}
\gamma(t) = \Theta(t) \frac{2}{M\pi} \int_0^{\infty}
{\rm d}\omega \frac{J(\omega)}{\omega} \cos(\omega t),
\end{equation}
where $\Theta(t)$ is a step function.
The frequency-independent damping $\gamma(t)=\gamma \delta(t)$,
called the Ohmic damping,
is obtained by taking the spectral function as $J(\omega)
=M \gamma \omega$. 

In the framework of the Caldeira-Leggett theory,
various situations have been studied by changing the
form of the potential. For example, the simplest situation is
a case of a flat potential ($V(q)=0$). In this situation,
the particle shows the Brownian motion.~\cite{Hakim85,Grabert88} 
The diffusion of a particle has been studied by using the
real-time path integrals based on the Feynman-Vernon 
method,~\cite{Grabert88,Feynman63} while the transport properties
of a dissipative particle can be described 
by the simple equation of motion
\begin{equation}
M \langle \ddot q(t) \rangle
 + M \int_{-\infty}^{t} {\rm d}t'
\, \gamma(t-t') \langle \dot q(t') \rangle
 = F(t),
\label{Brown1}
\end{equation}
for the average position $\langle q(t) \rangle$. The random 
force does not affect the response of the external force $F(t)$,
because $\langle \xi(t) \rangle = 0$. Particularly for the ohmic damping
case $\gamma(t) = \gamma \delta(t)$, 
the Caldeira-Leggett theory corresponds to the naive
Drude theory, where $\gamma^{-1}$ is the average collision time.

On the other hand, the Caldeira-Leggett model has
been used to study how the quantum coherence is suppressed
by dissipation in the double well 
potential,~\cite{Leggett87} which can be truncated
to two-state systems called the spin-boson model.
It has been proved that the spin-boson model
shows various behaviors 
including coherent motion at low temperatures for weak damping,
and incoherent motion in the limit of high temperatures and/or strong 
damping. This feature has been obtained first by an analytical
approximation~\cite{Leggett87}, and recently by numerical
simulations.~\cite{Chakravarty95,Costi96}
The spin-boson model with ohmic damping
can be related to the Kondo model~\cite{Kondo64} 
by the bosonization technique.~\cite{Guinea85} Moreover,
the spin-boson model is relevant to the Fermi edge 
singularities~\cite{Kondo88} which appears in 
the X-ray absorption edge anomaly~\cite{Nozieres69} and the diffusion 
of a heavy particle in metals.~\cite{Kondo76,Kondo84,Yamada84,
Kadono89,Luke91} Though these problems have been solved first by other
methods, it has been shown that the spin-boson model
gives the same result for them.~\cite{Chakravarty85,Sols87}
Recently, it has been claimed that
the spin-boson model with ohmic damping 
is also relevant to $c$-axis transport in high-$T_{\rm c}$ 
superconductors ~\cite{Clarke95,Clarke97} and 
a quantum dot coupled to an additional quantum point 
contact.~\cite{Aleiner97}

In this paper, we consider dissipative dynamics
of a quantum particle in the tight-binding model
through thermodynamics and transport properties.
This model for ohmic damping is relevant, 
with certain qualification, to the dissipative motion of small 
Josephson junctions~\cite{Zwerger87,Schoen90},
vortices in long Josephson junctions~\cite{Kato96} and 
heavy particles in a metal.~\cite{Sols87,Hedegard87a,
Hedegard87b} We take this model as a phenomenological
model to describe the quantum transport affected by
dissipation, which arises from coupling to
other degrees of freedom. In fact, this phenomenological 
introduction is generally justified at least in the high-temperature
limit.~\cite{Hedegard87a,footnote1} 
We focus mainly on the weak-damping region. In this paper,
the main aim is to study the crossover behavior from the incoherent
motion at high temperatures to the quantum motion 
at low temperatures. At high temperatures,
the particle motion is described
by the classical master equation in which the 
tunneling rate is given by the same value 
as two-state systems,~\cite{Weiss85}
and the optical conductivity shows an unusual $\omega$-dependence
because of the Fermi edge singularities as seen in two-state systems.
On the other hand, in the continuum limit which corresponds to
the low-temperature and weak-damping limit, we recover
the Drude form derived by the equation of motion (\ref{Brown1}) with 
the mass $M$ replaced by an effective mass $\Mbar$,
where $\Mbar$ is a curvature at the bottom of the particle band.
We give the exact formulation
for the crossover behavior in the weak-coupling region,
and connect the two different regions.
We also speculate that this study may
contribute to understand non-Drude forms of the conductivity
observed in various strongly correlated systems.
(See \S~\ref{OhmicGreen}.)

\begin{figure}[tb]
\hfil
\epsfile{file=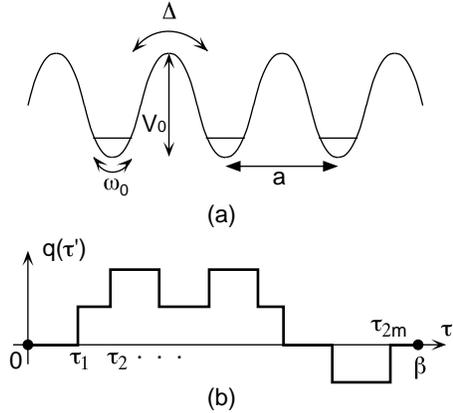,scale=0.5}
\hfil
\caption{(a) The periodic potential studied in this paper,
and (b) a representative path in the imaginary-time path integral 
in (\ref{ZR}).}
\label{pot}
\end{figure}

Here, we introduce the model and review the previous works
on this model. In this paper, we consider a particle of mass $M$ 
moving in a periodic potential $V(q+a)=V(q)$
as shown in Fig.~\ref{pot}~(a).
Here, a barrier height is denoted by $V_0$, and
a frequency of small oscillation around potential minima 
by $\omega_0$. We assume $V_0 \gg \hbar\omega_0 \gg k_B T$, where
$k_B$ is the Boltzmann constant, and $T$ is a temperature.
In this parameter region, the system can be reduced to
a tight binding model, which has been studied by
several authors in the framework of the real-time path
integral.~\cite{Weiss85,Weiss88,Chen89,Weiss91,Sassetti92} 
Fisher and Zwerger 
have studied dynamics of a particle in a periodic potential
in the weak corrugation limit ($V_0\rightarrow 0$), and
have related the results to the tight-binding model
through the duality transformation,~\cite{Fisher85,Schmid83}
which is also used in the subsequent
works.~\cite{Zwerger87,Eckern87} 
In these works, one of the important quantities 
is the linear mobility $\mu_l$ of the particle defined 
in the long-time limit $t\rightarrow\infty$ as
\begin{equation}
   \mu_l = \lim_{F \rightarrow 0} \lim_{t \rightarrow \infty}
   \frac{\langle q(t) \rangle}{Ft},
\end{equation}
where $F$ is an external force on a particle. 
The temperature dependence of $\mu_l$, however, has been exactly
calculated only for a special value of dimensionless damping 
strength, $K=1/2$, where $K=M \gamma
a^2/2\pi \hbar$.~\cite{Sassetti92}

In this work, we focus mainly on a weak coupling region $K\ll 1$,
and calculate specific heat and optical conductivity $\sigma(\omega)$.
The linear mobility is
also obtained by $\mu_l = \sigma(\omega\rightarrow 0)/e^2$.
Thermodynamics of the system is formulated
by the imaginary-time path integral,
and transport properties are studied in the framework
of the linear response theory. 

The paper is organized as follows. In \S~\ref{Formulation},
we derive exact formal expressions for the partition function
and the optical conductivity.
We also formulate a continuous limit,
an incoherent limit and a weak coupling theory for arbitrary form
of $J(\omega)$. For analytical calculations, $J(\omega)$ is 
taken as
\begin{equation}
J(\omega) = \frac{2\pi \delta_s}{a^2} 
\left(\frac{\omega}{\widetilde{\omega}}
\right)^{s-1} \omega f(\omega;\omega_c),
\label{Jomegad}
\end{equation}
where $\delta_s$ is a dimensionless coupling coefficient, and
$\widetilde{\omega}$ is a reference frequency.
Here, $f(\omega;\omega_c)$ is a cutoff function usually taken
as $f(\omega, \omega_c) = {\rm e}^{-\omega/\omega_c}$, while
we use a sharp cutoff,
$f(\omega,\omega_c) = \Theta(\omega_c - \omega)$
in \S~\ref{Continuous},
where $\Theta(\omega)$ is a step function.
In any case, the low-frequency 
($\omega\ll \omega_c$) behaviors of the dissipative particle
are not influenced by the detailed form of the cut-off function.
The exponent $s$ in (\ref{Jomegad}) mainly determines 
the properties of the heat bath. 
In \S~\ref{Ohmic}, we study the case $s=1$  
called `ohmic damping', where the coupling constant
$\delta_s$ agrees with $K$. For the ohmic case,
application of this calculation to real systems 
is also discussed in \S~\ref{OhmicGreen}.
In the case $s>1$ called `superohmic damping', and in the case $s<1$
called `subohmic damping', calculations are so
complicated that we only focus on the continuous limit 
(\S~\ref{Continuous}) and the incoherent regime
(\S~\ref{fhigh}). 
Summary is given in \S~\ref{Summary}.
Throughout this paper, we take $\hbar=k_B=1$.

\section{Formulation}\label{Formulation}
In this section, we give formal expressions for
the partition function (\S~\ref{fpart}), the optical conductivity 
(\S~\ref{fopt}).
We also discuss a continuous limit of the
tight binding model (\S~\ref{Continuous}). 
To deal with this model analytically, we focus on
two limiting regions: an incoherent tunneling regime
(\S~\ref{fhigh}) and a weak damping region (\S~\ref{weak}).
In this section, the derived expressions are applicable to
arbitrary form of $J(\omega)$.

\subsection{Partition function}\label{fpart}
The exact formal expression for the partition function $Z$
is given in the imaginary-time functional-integral representation.
The integrals over the heat-bath variables are Gaussian and they
can be evaluated exactly.~\cite{Weiss93,Feynman72}
After integrating out the degrees of freedom
for the environment, the partition function is expressed as
\begin{equation}
Z = Z_R \oint {\cal D} q(\tau) \exp(- S[q(\tau)]).
\label{ZR}
\end{equation}
Here, $Z_R$ is the partition function of the heat bath,
and in following calculation, $Z_R$ is removed
in order to focus only on the damping
effects on the particle. 
The paths $q(\tau)$ satisfy the periodic boundary condition
$q(\beta)=q(0)$, where $\beta = 1/T$.
The effective action $S[q(\tau)]$ is given by
\begin{equation}
S[q(\tau)] = \int_0^{\beta} {\rm d}\tau \left( \frac12 M \dot q^2
+V(q) \right) 
- \frac{1}{a^2} \int_0^{\beta} {\rm d}\tau {\rm d}\tau' \,
\phi(\tau-\tau') 
\dot q(\tau) \dot q(\tau').
\label{Seff}
\end{equation}
Damping effects due to the heat bath are described 
by the last term, which takes a nonlocal form for $\tau$.
The kernel $\phi(\tau)$ is calculated as
\begin{eqnarray}
\phi(\tau) &=& \frac{a^2}{\pi} \int_0^{\infty} {\rm d}\omega
\frac{J(\omega)}{\omega^2} (D_{\omega}(0)-D_{\omega}(\tau)), 
\label{phidef} \\
D_{\omega}(\tau) &=& \frac{1}{\beta} \sum_{n=-\infty}^{\infty}
\frac{2\omega}{\nu_n^2+\omega^2}{\rm e}^{{\rm i}\nu_n \tau} 
\label{Domega} \\
&=& \frac{\cosh[\omega(\beta/2-|\tau|)]}{\sinh(\beta \omega/2)},
\label{Domega2}
\end{eqnarray}
where $\nu_n = 2\pi n/\beta$ is the Matsubara frequency. 

The form of the potential $V(q)$ is shown in Fig.~\ref{pot} (a).
In the parameter region $T \ll \omega_0 \ll V_0$,
the system is effectively mapped to a tight-binding model. In this case, 
the value of paths $q(\tau)$ takes $q = a n$ for almost all $\tau$, 
where $n$ is an integer.
A representative path is shown in Fig.~\ref{pot} (b).
Though the truncation scheme for the paths has already been studied 
in the literature,~\cite{Chakravarty85,Dorsey86,Weiss87} 
we will describe it briefly for self-contained description.

In the dissipationless case, a bare hopping matrix $\Delta_0$ 
is determined by the instanton action $S_0$ as
\begin{equation}
   \Delta_0 = A \exp( - S_0),
\end{equation}
where $A$ is a prefactor of the order of $\omega_0$.~\cite{Callan77}
Hence, we have a tight binding model described by the Hamiltonian
\begin{equation}
   H = - \Delta_0 \sum_n \left( c^{\dagger}_{n+1} c_n +
   c^{\dagger}_n c_{n+1} \right) ,
\end{equation}
where $c_n (c^{\dagger}_n)$ are annihilation (creation)
operators of a spinless fermion at site $n$.
The damping effects due to the heat bath 
can be divided into the fast and slow modes as 
\begin{eqnarray}
   J(\omega) &=&  J_{\rm lf}(\omega) + 
   J_{\rm hf}(\omega)  \\
   J_{\rm lf}(\omega) &=& J(\omega) f(\omega;\omega_c)
   \label{spec} 
\end{eqnarray}
Here, $\omega_c$ is a cutoff frequency ($\Delta_0, T
\ll \omega_c \ll \omega_0$), and $f(\omega;\omega_c)$
is a cut-off function taken as $f(\omega;\omega_c)\ll 1$ 
for $\omega_c \ll \omega$.
Then, the fast modes can be treated easily by the Born-Oppenheimer 
approximation. In the approximation, the bare 
matrix element $\Delta_0$ is reduced to a renormalized 
one~\cite{Chakravarty82,Bray82}
\begin{equation}
   \Delta = \Delta_0 \exp \left( - \frac{a^2}{2\pi} 
   \int_0^{\infty} {\rm d}\omega \frac{J_{\rm hf}(\omega)}{\omega^2} 
   \right) .
\end{equation}
In contrast to the fast modes, the influence of the slow modes
of the heat bath must be studied carefully
based on the effective action (\ref{Seff}). 
In this paper, $J_{\rm lf}(\omega)$ is replaced with $J(\omega)$,
and $\Delta$ and $\omega_c$ are regarded
as free parameters ($\Delta, T \ll \omega_c$).

Because we have already traced out the fast modes which can follow the
instanton motion of the time scale $1/\omega_0$, 
each path $\dot q(\tau)$ is
expressed as the sum of the delta functions as
\begin{equation}
   \dot q(\tau)/a = \sum_{l=1}^{2m} \xi_l \delta(\tau-\tau_l).
   \label{dotq}
\end{equation}
Here, $\xi_l=\pm 1$ specifies the direction of the $l$-th hopping
at $\tau=\tau_l$. Substituting (\ref{dotq}) to (\ref{Seff}), 
the exact formal expression of the partition function
is obtained as
\begin{eqnarray}
   Z &=& \sum_{m=0}^{\infty} \Delta^{2m} \sum_{ \{ \xi_l \}'}
   \int_0^{\beta} {\rm d}\tau_{2m} \int_0^{\tau_{2m}} {\rm d}\tau_{2m-1}
   \cdots \int_0^{\tau_2} {\rm d}\tau_1 
   \exp\left[ \sum_{k<l}^{2m} 
   \xi_k \xi_l \phi(\tau_l - \tau_k) \right] \nonumber \\
   &=& \sum_{m=0}^{\infty} \frac{\Delta^{2m}}{2m!} 
   \sum_{\{\xi_l\}'}
   \prod_{n=1}^{2m} \int_0^{\beta} {\rm d}\tau_n 
   \exp\left[ \sum_{k<l}^{2m} \xi_k \xi_l \phi(\tau_l-\tau_k) \right] .
   \label{clZ}
\end{eqnarray}
Here, the prime in $\{\xi_l\}'$ denotes summation in
accordance with the constraint
\begin{equation}
   \sum_{l=1}^{2m} \xi_l = 0.
   \label{constraint}
\end{equation}

The expression (\ref{clZ}) may be interpreted as
the statistical model of 
the classical particles interacting to the $\tau$-direction 
with the potential $-\phi(\tau)$.~\cite{Schmid83} 
In this point of view, $\Delta$ and $\xi_l$ are regarded 
as a chemical potential and a charge of the classical particle.
The inverse temperature $\beta=1/T$ in the original model
corresponds to the system length in the $\tau$ direction.
At the same time,
the constraint (\ref{constraint}) is interpreted as
the condition for the charge neutrality. 
The mapping to the classical statistical mechanics is useful for
constituting the weak coupling theory. Actually, the 
`screening effect' by the charged particles is important 
to describe transport properties of the original model
at low temperatures. (See \S~\ref{weak}.)

\subsection{Optical conductivity}\label{fopt}
We formulate the optical conductivity $\sigma(\omega)$ 
in one dimension using the Kubo formula~\cite{Kubo57}
\begin{equation}
\sigma(\omega) = -e^2 a^2\frac{\langle -{\cal K} \rangle 
- \Lambda(\omega)} {{\rm i}(\omega + {\rm i} \delta)}, 
\label{Kubo}
\end{equation}
where $\delta$ is an adiabatic constant, and
${\cal K}$ is a kinetic energy
\begin{equation}
{\cal K} = - \Delta \sum_n \left( c^{\dagger}_{n+1} c_n +
c^{\dagger}_{n} c_{n+1} \right), 
\end{equation}
and a current-current correlation function
$\Lambda(\omega)$ is defined by
\begin{eqnarray}
\Lambda(\omega) &=& \widetilde{\Lambda}({\rm i}\omega_m \rightarrow
\omega + {\rm i}\delta). 
\label{realcorr} \\
\widetilde{\Lambda}({\rm i}\omega_m) &=& \int_0^{\beta} {\rm d}\tau
\, {\rm e}^{{\rm i}\omega_m \tau} \langle j(\tau) j(0) \rangle, 
\end{eqnarray}
Here, the current operator $j(\tau) = {\rm e}^{\tau H} j 
{\rm e}^{-\tau H}$ 
is defined by
\begin{equation}
j = {\rm i} \Delta \sum_{n} \left( c^{\dagger}_{n+1} c_n -
c^{\dagger}_{n} c_{n+1}\right).
\label{current}
\end{equation}
The real part of $\sigma(\omega)$
contains a coherent part expressed by
a delta function as~\cite{Scalapino93,Kohn64}
\begin{equation}
{\rm Re} \,
\sigma(\omega) = D \delta(\omega) + \sigma_{\rm res}(\omega).
\label{Drudeweight} 
\end{equation}
Here, $D$ is called a Drude weight, and $\sigma_{\rm res}(\omega)$ 
is a residual incoherent part.
From (\ref{Kubo}), we obtain
\begin{eqnarray}
& & \frac{D}{\pi e^2 a^2} = \langle -{\cal K} \rangle - {\rm Re} 
\, \Lambda(\omega \rightarrow 0), \\
& & \sigma_{\rm res}(\omega) = e^2 a^2 
\frac{{\rm Im} \, \Lambda(\omega)}{\omega}.
\label{regres}
\end{eqnarray}

To evaluate $\Lambda(\omega)$, we consider the imaginary-time
correlation function $\widetilde{\Lambda}(\tau) = 
\langle j(\tau) j(0) \rangle$ for $\tau>0$.
The correlation function is formulated by
the imaginary-time path integral as
\begin{equation}
  \widetilde{\Lambda}(\tau)
= \frac{\Delta^2}{Z} \oint {\cal D}' q(\tau')
  \exp \left[-S[q(\tau')] \right].
\label{Lambdaeff}
\end{equation}
Here, the path integral is performed over all the possible
paths $q(\tau)$ with two jumps at $\tau'=0, \tau$ as 
\begin{equation}
\dot q(\tau') / a = \sum_{l=1}^{2m} \xi_l \delta(\tau'-\tau_l) 
+ \sigma \delta(\tau') + \sigma' \delta(\tau'-\tau),
\label{qdot2}
\end{equation}
where $2m+2$ is the number of transitions,
and $\xi_l, \sigma, \sigma' = \pm 1$ denote the directions of
hopping at $\tau' = \tau_l, 0, \tau$, respectively.  
Then, the path integral is expressed as
\begin{equation}
\oint {\cal D}' q(\tau') (\cdots) = \sum_{m=0}^{\infty}
\sum_{\{ \xi_l, \sigma, \sigma' \}'} (-\sigma \sigma')
\frac{1}{2m!} \prod_{l=1}^{2m} \int_0^{\beta} {\rm d}\tau_l
(\cdots ).
\label{integ2}
\end{equation}
Here, the prime in $\{\xi_l, \sigma, \sigma' \}'$ 
denotes the summation in accordance with the constraint
\begin{equation}
\sum_{l=1}^{2m} \xi_l + \sigma + \sigma' = 0,
\end{equation}
which comes from the periodic boundary condition for $q(\tau')$.
Here, we divide $\Lambda(\omega)$ into two parts as
\begin{equation}
\widetilde{\Lambda}(\tau) = \widetilde{\Lambda}_1(\tau)
-  \widetilde{\Lambda}_2(\tau),
\label{lambda0}
\end{equation}
where $\widetilde{\Lambda}_1(\tau)$ and $\widetilde{\Lambda}_2(\tau)$
denote the contribution of paths which satisfy 
$\sigma = -\sigma'$ and $\sigma = \sigma'$, respectively.
From (\ref{Lambdaeff})-(\ref{integ2}), we obtain
\begin{eqnarray}
\widetilde{\Lambda}_1(\tau) &=& \frac{2\Delta^2}{Z} \sum_{m=0}^{\infty}
\frac{\Delta^{2m}}{2m!} \sum_{\{\xi_l\}'} \int_0^{\beta} 
\prod_{l=1}^{2m} {\rm d}\tau_l \nonumber \\
& & \times \exp \left( \sum_{k<l}^{2m}
\xi_k \xi_l \phi(\tau_l-\tau_k) + \sum_{l=1}^{2m} \xi_l \phi(\tau_l)
-\sum_{l=1}^{2m} \xi_l \phi(\tau_l-\tau) - \phi(\tau) \right). 
\label{lambda1} \\
\widetilde{\Lambda}_2(\tau) &=& \frac{2\Delta^2}{Z} \sum_{m=0}^{\infty}
\frac{\Delta^{2m}}{2m!} \sum_{\{\xi_l\}''} \int_0^{\beta} 
\prod_{l=1}^{2m} {\rm d}\tau_l \nonumber \\
& & \exp \left( \sum_{k<l}^{2m}
\xi_k \xi_l \phi(\tau_l-\tau_k) + \sum_{l=1}^{2m} \xi_l \phi(\tau_l)
+\sum_{l=1}^{2m} \xi_l \phi(\tau_l-\tau) + \phi(\tau) \right).
\label{lambda2}
\end{eqnarray}
Here, the prime in $\{\xi_l\}'$ and
the double prime in $\{\xi_l\}''$ denote the summations
in accordance with (\ref{constraint}) and the condition
\begin{equation}
\sum_{l=1}^{2m} \xi_l + 2 = 0,
\label{constraint2}
\end{equation}
respectively. From (\ref{lambda1}) and (\ref{lambda2}),
$\widetilde{\Lambda}_1(\tau)$ and 
$\widetilde{\Lambda}_2(\tau)$ can be interpreted as
the partition functions with fixed charges at $\tau'=0, \tau$
in terms of the classical partition function (\ref{clZ}).

We can calculate $\sigma_{\rm res}(\omega)=e^2 a^2 {\rm Im}
\Lambda(\omega)/\omega$ without the Wick rotation (\ref{realcorr}).
We introduce a real-time correlation function as
\begin{eqnarray}
\Lambda(t) &=& {\rm i} \Theta(t) 
\langle j(t) j(0) - j(0) j(t)  \rangle \nonumber \\
&=& -2{\rm Im}\, \widetilde{\Lambda}(\tau\rightarrow {\rm i}t),
\label{realform1}
\end{eqnarray}
where $j(t)={\rm e}^{{\rm i}Ht}
j {\rm e}^{-{\rm i}Ht}$. Then, the correlation function $\Lambda(\omega)$
is calculated by
\begin{equation}
\Lambda(\omega) = \int_0^{\infty} {\rm d}t 
\, \Lambda(t) {\rm e}^{{\rm i}
\omega t}. \label{LambdaFourier}
\label{realform2}
\end{equation} 
The above expressions, (\ref{realform1}) and (\ref{realform2}),
are used to the calculation
in the incoherent region and the weak-coupling region.
We should note that the causality, 
$\Lambda(t)=0$ for $t<0$, gives 
the frequency sum-rule~\cite{Shastry90} 
\begin{equation} 
\int_{-\infty}^{\infty} {\rm d}\omega 
\, {\rm Re} \, \sigma(\omega) = \pi e^2 a^2\langle 
-{\cal K} \rangle.
\label{sumrule}
\end{equation}
The sum rule is always satisfied in the approximations
adopted in this paper, since the causality is always satisfied.

\subsection{Continuum limit}\label{Continuous}
In the dissipationless case, the energy dispersion 
is given by
\begin{equation}
\eps_k = - 2 \Delta \cos ka,
\end{equation}
where $k$ is a momentum. 
The lattice model is reduced to the continuum model
in the limit $a \rightarrow 0$ ($\Delta\rightarrow \infty$)
by keeping $\Mbar=1/2 \Delta a^2$ constant. 
In this limit, the energy dispersion is approximated as
\begin{equation}
  \eps_k \simeq -2\Delta + \frac{k^2}{2\Mbar}.
\end{equation} 
Note that the mass $\Mbar$ is not necessarily the same as
the bare mass $M$ in the original
Hamiltonian. In the continuum limit $\Delta \rightarrow \infty$,
thermal fluctuations may be neglected 
compared with the band width ($T/\Delta\rightarrow 0$).
Hence, the continuum limit corresponds to the
zero-temperature limit.

Next, we consider the dissipative case.
In the continuum limit, the classical equation of motion for the
average position $\langle q(t) \rangle$ is obtained by virtue of
Ehrenfest's theorem as
\begin{equation}
  \langle \ddot q(t) \rangle + \int_{-\infty}^{t} dt' \gamma(t-t')
  \langle \dot q(t') \rangle = \frac{e}{\Mbar} E(t),
\label{eqofmotion}
\end{equation}
where $E(t)$ is an external electric field.  The damping kernel
$\gamma(t)$ is determined by the spectral density $J(\omega)$
as~\cite{Weiss93,Grabert88}
\begin{eqnarray}
  \gamma(t) &=& \int_{-\infty}^{\infty} \frac{{\rm d}\omega}{2\pi}
  {\rm e}^{-{\rm i}\omega t}
  \widehat{\gamma}(z\rightarrow -{\rm i} \omega +\delta), \\
  \widehat{\gamma} (z) &=& \frac{2z}{\pi \Mbar}
  \int_0^{\infty} {\rm d}\omega' 
  \frac{J(\omega')}{\omega' (\omega'^2 + z^2)}.
  \label{gammaz}
\end{eqnarray}
From the classical equation (\ref{eqofmotion}), 
the Fourier transformation of the current
$j(t)=e\langle \dot q(t) \rangle$ gives
$\widetilde j(\omega) = \sigma(\omega) \widetilde E(\omega)$,
and the optical conductivity $\sigma(\omega)$ is given by
\begin{equation}
  \sigma{(\omega)} = \left. \frac{e^2}{\Mbar
(z+\widehat{\gamma}(z) )} 
\right|_{z\rightarrow -{\rm i}\omega + \delta} .
\label{contres}
\end{equation}
Because of the causality,
the sum rule (\ref{sumrule}) is also satisfied in the continuum limit.

The above result is valid for any form of 
the spectral density $J(\omega)$.
Here, we take $J(\omega)$ as
\begin{equation}
J(\omega)= \Mbar\gamma_s \left(\frac{\omega}{\widetilde{\omega}}
\right)^{s-1} \omega \Theta(\omega_c -\omega),
\label{Jomega}
\end{equation}
where $\gamma_s = 2\pi\delta_s/a^2 \Mbar 
= 4\pi\delta_s\Delta$ is a damping frequency,
and the cut-off function $f(\omega;\omega_c)$
in (\ref{Jomegad}) is taken as 
the step function $\Theta(\omega_c-\omega)$ for convenience.
The characteristic damping frequency
$\widetilde{\gamma}$ is determined by $\widetilde{\gamma}
=(\gamma_s \widetilde{\omega}^{1-s})^{1/(2-s)}$.
To keep $\widetilde{\gamma}$ constant
in the continuum limit $\Delta\rightarrow \infty$,
the dimensionless coupling coefficient $\delta_s
= \gamma_s/4\pi \Delta$ must be suppressed to zero.
Hence, the continuum limit
corresponds to the weak coupling limit $\delta_s\rightarrow 0$.

From (\ref{Jomega}), the damping
kernel $\widehat{\gamma}(\omega)$
in (\ref{gammaz}) is calculated analytically.
The leading term is given by~\cite{Grabert88}
\begin{equation}
\widehat{\gamma}(z)=\left\{
\begin{array}{ll}
\dis{\frac{\gamma_s}{\sin(\pi s/2)} \left(\frac{z}{\widetilde{\omega}}
\right)^{s-1}
\left[ 1+ {\cal O}\left( z/\omega_c, (z/\omega_c)^{2-s} 
\right) \right]}, & (0<s<2), \\
\dis{\frac{\gamma_s z}{\pi \widetilde{\omega}} \ln\left(1+
\frac{\omega_c^2}{z^2}\right)}, & (s=2), \\
\dis{\frac{2\gamma_s}{\pi(s-2)}\left(
\frac{\omega_c}{\widetilde{\omega}}\right)^{s-2}
\frac{z}{\widetilde{\omega}} \left( 1+
{\cal O}(z^2/\omega_c^2, (z/\omega_c)^{s-2}) \right)}, & (s>2).
\end{array}
\right.
\label{gamma}
\end{equation}
From (\ref{contres}) and (\ref{gamma}),
it is proved that
the delta function disappears at $\omega= 0$ 
in ${\rm Re}\,\sigma(\omega)$ for $0<s<2$.
This result strongly indicates that the Drude weight $D$ 
in (\ref{Drudeweight}) vanishes for $0<s<2$ at
all temperatures and damping coefficients.
This is because the continuum limit describes
the low-temperature and weak-coupling limit, 
$T, \delta_s \rightarrow 0$, while this limit should give
the most coherent result in the parameter space.
On the other hand, the Drude weight exists for $s>2$, and
the damping effects appears only in the mass renormalization
\begin{equation}
\Mbar \rightarrow \Mbar \left(1+\frac{2}{\pi(s-2)}
\frac{\gamma_s}{\widetilde{\omega}}
\left(\frac{\omega_c}{\widetilde{\omega}}\right)^{s-2}\right).
\end{equation}
In this case, the residual conductivity for $\omega>0$ vanishes.
The above result is related to the Brownian motion caused by
a heat bath, where the long time
behavior of $\langle (q(t)-q(0))^2 \rangle$ 
at finite temperatures is proportional
to $t^2$ for $s>2$, and $t^s$ for $s<2$.
The transition at $s=2$ can be understood by a simple 
discussion on the mass renormalization
expressed in general as~\cite{Hakim85,Grabert88} 
\begin{equation}
\Mbar \rightarrow \Mbar + \frac{2}{\pi} \int_0^{\infty}
{\rm d}\omega \frac{J(\omega)}{\omega^3}.
\end{equation}
When this integral is convergent, the particle motion becomes
coherent in the continuum limit.

\begin{figure}[tb]
\hfil
\epsfile{file=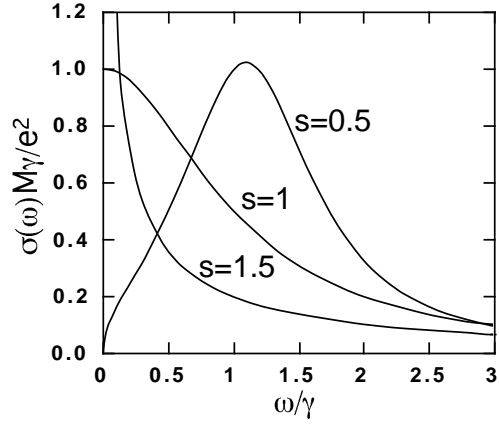,scale=0.65}
\hfil
\caption{The optical conductivity 
$\sigma(\omega)$ in the continuum limit for $s=0.5, 1, 1.5$.
Particularly, for the ohmic damping $s=1$, $\sigma(\omega)$ 
takes a simple Drude form.}
\label{contgraph}
\end{figure}

Let us see the $s$ dependence in more detail.
The residual part $\sigma_{\rm res}(\omega)$ 
appears only for the case $s\le 2$.
From (\ref{contres}) and (\ref{gamma}),
we obtain
\begin{equation}
\sigma_{\rm res}(\omega) =\frac{e^2}{\Mbar \widetilde{\gamma}}
\frac{\dis{\left(\frac{\omega}{\widetilde{\gamma}}\right)^{s-1}}}
{\dis{\left(\frac{\omega}{\widetilde{\gamma}}
-\cot \frac{\pi s}{2}\left(\frac{\omega}{\widetilde{\gamma}}
\right)^{s-1} \right)^2 + \left(\frac{\omega}{\widetilde{\gamma}}
\right)^{2(s-1)}}},
\label{contopteq}
\end{equation}
where $\widetilde{\gamma} = \left(\gamma_s\widetilde{\omega}^{1-s}
\right)^{1/(2-s)}$. In this case,
$\sigma_{\rm res}(\omega)$ behaves qualitatively as
\begin{equation}
\sigma_{\rm res}(\omega) \sim
\left\{ \begin{array}{ll} \omega^{1-s} & (\omega < \widetilde{\gamma}),
\\ \omega^{s-3} & (\omega >\widetilde{\gamma}). \end{array} \right.
\label{contqb}
\end{equation}
The $\omega$-dependence of $\sigma_{\rm res}(\omega)$ 
for $s=0.5, 1, 1.5$ is shown in Fig.~\ref{contgraph}.
For subohmic damping $0<s<1$, the conductivity
is reduced to zero as $\omega$ decreases. On the other hand,
in the superohmic case $1<s<2$, the conductivity diverges 
for $\omega\rightarrow 0$. 
For the ohmic damping $s=1$, the optical conductivity
$\sigma_{\rm res}(\omega)$ takes a simple Drude form.

As $s$ increases toward $2$,
the curve of $\sigma_{\rm res}(\omega)$ approaches the 
$1/\omega$ form. 
The case of $s=2$ is marginal, and we obtain
\begin{equation}
\sigma_{\rm res}(\omega) = 
\frac{e^2\gamma_2}{\Mbar\omega\widetilde{\omega}} 
\frac{1}{\dis{
\left(1+\frac{\gamma_2}{\pi \widetilde{\omega}}
\log \bigr|\frac{\omega_c^2}{\omega^2}-1\bigl| \right)^2 
+\left(\frac{\gamma_2}{\widetilde{\omega}}\right)^2
}}, \hspace{5mm} (\omega<\omega_c) .
\end{equation}
The optical conductivity $\sigma_{\rm res}
(\omega)$ behaves nearly as $1/\omega$ for $\omega\ll\omega_c$.
For $s>2$, the incoherent part vanishes, and
the delta function appears in ${\rm Re} \,\sigma(\omega)$.

\subsection{Incoherent tunneling regime}\label{fhigh}
In the limit of high temperatures and/or strong damping, 
the particle moves incoherently to the neighbor sites.
In this regime, the occupation probabilities obey a simple
master equation,~\cite{Fisher85,Weiss85}
and the hopping rate is formulated by the Fermi's golden rule,
where the incoherent motion of the particle is described 
by the term of order of $\Delta^2$ in the correlation function.
From (\ref{lambda0})-(\ref{lambda2}),
the relevant term is given by
\begin{equation}
\widetilde{\Lambda}(\tau)=2\Delta^2\exp\left[-\phi(\tau)\right].
\label{hightemp}
\end{equation}
From (\ref{realform1}), the real-time correlation 
function $\Lambda(t)$ is derived as
\begin{equation}
\Lambda(t) =  4\Delta^2 \sin R(t) \exp(-S(t)),
\label{highrealcorr}
\end{equation}
where $S(t)$ and $R(t)$ are real functions defined by
$\phi(\tau={\rm i}t) = S(t) + {\rm i}R(t)$. The explicit form
of $S(t)$ and $R(t)$ is given from (\ref{phidef}) and 
(\ref{Domega2}) as
\begin{eqnarray}
S(t) &=& \frac{a^2}{\pi} \int_0^{\infty} {\rm d}\omega
\frac{J(\omega)}{\omega^2} (1-\cos \omega t) \coth 
\frac{\beta \omega}{2}, 
\label{Stdef} \\
R(t) &=& \frac{a^2}{\pi} \int_0^{\infty} {\rm d}\omega
\frac{J(\omega)}{\omega^2} \sin\omega t.
\label{Rtdef}
\end{eqnarray}
The Fourier transformation of (\ref{highrealcorr}) gives
\begin{equation}
{\rm Im} \, \Lambda(\omega) = 4\Delta^2 \int_0^{\infty} 
{\rm d}t \, \sin \omega t
\sin R(t) \exp( -S(t) ).
\label{highcorr2}
\end{equation}
Thus, the optical conductivity is calculated from $\sigma_{\rm res}
(\omega)=e^2a^2 {\rm Im}\,\Lambda(\omega)/\omega$. 
The formulation of the incoherent region is valid
only when the integral (\ref{highcorr2})
is convergent. 

The above formulation is quite similar to the calculation of
the nonlinear mobility in biased periodic-potential 
systems.~\cite{Weiss91}
Actually, the tunneling rate $\Gamma(\eps)$ and 
the nonlinear mobility $\mu(\eps)$ with a bias $\eps$ 
are calculated in these systems as
\begin{eqnarray}
& & \Gamma(\eps) = 4\Delta^2 \int_0^{\infty} {\rm d}t
\cos \eps t \sin R(t) \exp(-S(t)), 
\label{incG} \\
& & \mu(\eps) = \frac{4a^2\Delta^2}{\eps}
\int_0^{\infty} {\rm d}t
\sin \eps t \sin R(t) \exp(-S(t)). 
\label{incmu}
\end{eqnarray}
Compared (\ref{incmu}) with (\ref{highcorr2}), 
the optical conductivity is obtained as $\sigma_{\rm res}
(\omega) = e^2 \mu(\eps\rightarrow \omega)$.
Further, it is proved that $\mu(\eps)$ is related
to $\Gamma(\eps)$ through the relation~\cite{Weiss91}
\begin{equation}
\mu(\eps) = a^2 \tanh(\beta\eps/2) \Gamma(\eps)/\eps.
\label{Gicrelation}
\end{equation}
Thus, we can obtain the optical conductivity from the 
tunneling rate $\Gamma(\eps)$ as
\begin{equation}
\sigma_{\rm res}(\omega) = e^2 a^2 \frac{\tanh(\beta\omega/2)}{\omega}
\Gamma(\eps\rightarrow\omega).
\label{incoptform}
\end{equation}
In order to calculate $\Gamma(\eps)$ in the incoherent
regime, we can utilize the results on two-state systems
obtained in ref.~\citen{Leggett87},
where the tunneling rate is formulated by the same form as
(\ref{incG}), and is calculated
for a few regions in the parameter space.
For the ohmic damping,
it is proved that the expansion in terms of $\Delta$ gives 
the systematic high-temperature expansion. 
We consider the lowest contribution of order of $\Delta^2$
in \S~\ref{Ohmic}. 
In the subohmic case ($0<s<1$) and the superohmic case ($1<s<2$),
the calculation
is so complicated that analytical treatment is possible
only in the limiting cases. The results are 
given in Appendix~\ref{app0}, and
we do not discuss the other cases in this paper.
We note that the result (\ref{highcorr2}) is also related to
the transition rate of the photo-induced tunneling 
in the two-level systems.~\cite{Chakravarty85}

In the incoherent tunneling regime,
the partition function (\ref{clZ}) is also approximated as
\begin{equation}
Z = 1 + 2\beta \Delta^2\int_0^{\beta} {\rm d}\tau \exp
[-\phi(\tau)].
\label{Zhigh}
\end{equation}
This form corresponds to the partition function 
of dissipative two-state systems 
except a difference of a factor 2 
which comes from two possible transitions to neighbor
sites.~\cite{Goerlich88} 

\subsection{Weak coupling theory}\label{weak}
For weak damping ($\delta_s \ll 1$), analytical calculation is
possible for all temperatures. 
We, however, should be careful to constitute the weak coupling
theory, because we must deal with screening effects
of interacting classical particles described in (\ref{clZ}).
In order to treat the screening effects, 
we adopt the `ring approximation', 
which is proved to be equivalent to the Debye-H\"uckel theory.
Since this approximation 
is believed to be valid in the weak coupling region,
we expect that reliable results is obtained
in the weak damping region by this approximation.
In this subsection, we derive the partition function and
the optical conductivity based on the ring approximation. 
Since details of the calculation are straightforward,
but rather tedious, we show only the results here.
Details of the calculation
are given in Appendix~\ref{app1} and \ref{app2}. 

\subsubsection{Partition function}
First, we define a Fourier transformation of the potential
$\phi(\tau)$ as
\begin{equation}
\phi({\rm i}\omega_m) = \int_0^{\beta} {\rm d}\tau 
\, {\rm e}^{{\rm i} \omega_m \tau} \phi(\tau),
\label{form4}
\end{equation}
where $\omega_m=2\pi m/\beta$ is the Matsubara frequency.
By comparing (\ref{phidef})-(\ref{Domega}) with (\ref{gammaz}), 
$\phi({\rm i}\omega_m)$ can be related to the damping kernel 
$\widehat{\gamma}(z)$ defined in (\ref{gammaz}) for $\omega_m \ne 0$ as
\begin{equation}
\phi({\rm i}\omega_m)=
-\frac{\Mbar a^2}{\omega_m} \widehat{\gamma}(z=\omega_m).
\label{potshift}
\end{equation}
To avoid unphysical divergence,
we shift the potential $\phi(\tau)$ by 
the zero frequency component $\phi_0=\phi({\rm i}\omega_m=0)$ as
\begin{equation}
\phi(\tau) \rightarrow \phi(\tau) + \phi_0.
\label{potrenorm}
\end{equation}
By the potential shift, the transition amplitude $\Delta$
is renormalized as
\begin{equation}
\Dbar = \Delta \exp\left(-\frac12 \phi_0\right),
\label{Dbardef}
\end{equation}
where the factor $\exp (-\phi_0/2)$ corresponds to the 
Frank-Condon factor.
Details of calculation for $\phi_0$ is given in Appendix~\ref{app1}. 

From the approximate calculation given in Appendix~\ref{app2},
the analytical form of the partition function is obtained as
\begin{eqnarray}
Z &=& \int_0^{2\pi} \frac{{\rm d}\theta}{2\pi} \exp
\left[ U(n) \right] 
\label{forms} \\
U(n) &=& n + \int_0^n {\rm d}n\, Q(n) - nQ(n),
\label{Ucal}
\\
Q(n) &=& \frac{1}{\beta^2} \sum_{\omega_m > 0}
\frac{n\phi({\rm i}\omega_m)^2}{1-n\phi
({\rm i}\omega_m)/\beta},
\label{Qcal}
\end{eqnarray}
where $n=n(\theta)$ is determined by the equation
\begin{equation}
2\beta\Dbar\cos\theta = n {\rm e}^{-Q(n)}.
\label{ndef}
\end{equation}
From eqs.~(\ref{forms})-(\ref{ndef}), the partition function $Z$ is
calculated analytically at least
for the ohmic damping. (See \S~\ref{heat}.) 

\subsubsection{Optical conductivity}
The ring approximation is also applicable to
the conductivity $\sigma(\omega)$. When we define 
a screened potential $\vphi(\tau;\theta)$ as
\begin{equation}
\vphi(\tau;\theta) = \frac{1}{\beta} \sum_{{\rm i} \omega_m}
\frac{\phi({\rm i}\omega_m)}{1-n(\theta)\phi({\rm i}\omega_m)/\beta} 
{\rm e}^{-{\rm i}\omega_m \tau},
\label{screenpotdef}
\end{equation}
the correlation functions are expressed as
\begin{eqnarray}
\widetilde{\Lambda}(\tau) &=& 
\widetilde{\Lambda}_1(\tau) 
- \widetilde{\Lambda}_2(\tau) 
\label{weaklambda0} \\
\widetilde{\Lambda}_1(\tau) 
&=& \frac{1}{2Z} \int_0^{2\pi}
\frac{{\rm d}\theta}{2\pi} \frac{n^2}{\beta^2\cos^2 \theta}
\exp\left[U(\theta)-\vphi(\tau;\theta)
\right], 
\label{weaklambda1}
\\
\widetilde{\Lambda}_2(\tau) 
&=& \frac{1}{2Z} \int_0^{2\pi} 
\frac{{\rm d}\theta}{2\pi} \frac{n^2}{\beta^2\cos^2 \theta}
\exp\left[U(\theta)+2{\rm i}\theta +
\vphi(\tau;\theta)\right].
\label{weaklambda2}
\end{eqnarray}
After the replacement $\tau \rightarrow {\rm i}t$,
the screened potential is expressed by
\begin{equation}
\vphi({\rm i}t;\theta) = S(t;\theta) + {\rm i}R(t;\theta),
\end{equation}
where $S(t;\theta)$ is a real part, and $R(t;\theta)$ is an imaginary part.
Using this notation, the real-time correlation function in
(\ref{realform1}) is written by
\begin{equation}
\Lambda(t) = \frac{1}{Z} 
\int_{0}^{2\pi} \frac{{\rm d}\theta}{2\pi}
\frac{n^2}{\beta^2 \cos^2\theta}
{\rm e}^{U(\theta)}
\left[{\rm e}^{-S(t;\theta)}+{\rm e}^{S(t;\theta)}\cos 2\theta \right]
\sin R(t;\theta).
\label{weakformalism}
\end{equation}
Thus, the optical conductivity is calculated 
from (\ref{regres}) and (\ref{realform2}).

In the high-temperature limit ($\beta\rightarrow \infty$), 
the screened potential (\ref{screenpotdef})
agrees with the unscreened potential $\phi(\tau)$, because
the second term of the denominator in (\ref{screenpotdef})
is suppressed. Therefore, the expressions of the weak coupling 
region is consistently
connected to the formulation in the incoherent regime,
(\ref{hightemp}).          

In the limit $T, \delta_s \rightarrow 0$,
the weak coupling must correspond to the continuum limit.
To see the relation, it is convenient to study
the imaginary-time correlation function $\widetilde{\Lambda}(\tau)$.
In the continuum limit, 
the dimensionless coupling coefficient $\delta_s$
is reduced to zero. Therefore, the screened potential 
$\vphi(\tau)$ becomes small, and the approximation
\begin{equation}
\exp(\pm\vphi(\tau)) \simeq 1 \pm \vphi(\tau)
\end{equation}
is justified. From (\ref{weaklambda0})-(\ref{weaklambda2}),
the imaginary-time correlation function is calculated
approximately as
\begin{equation}
\widetilde{\Lambda}({\rm i}\omega_m) =
-\frac{1}{2Z} \int_0^{2\pi} \frac{{\rm d}\theta}{2\pi}
\frac{n^2}{\beta^2 \cos^2\theta} {\rm e}^{U(\theta)} 
\left(1+{\rm e}^{2{\rm i}\theta} \right) \frac{\phi({\rm i}\omega_m)}
{1-n(\theta)\phi({\rm i}\omega_m)/\beta}.
\label{nemui}
\end{equation}
The continuum limit ($\Delta\rightarrow\infty$) corresponds 
also to the low-temperature limit ($T/\Delta\rightarrow 0$),
where the contribution at $\theta=0$ is dominant
in the integral (\ref{nemui}), and 
$n$ is evaluated by $n=2\beta\Delta$ from (\ref{ndef}). 
Thus, we obtain
\begin{equation}
\widetilde{\Lambda}({\rm i}\omega_m) = \frac{1}{\Mbar a^2}
\frac{\hat{\gamma}(\omega_m)}{\omega_m + \hat{\gamma}(\omega_m)},
\label{nemui2}
\end{equation}
where $\Mbar =1/2\Delta a^2$ and the relation 
(\ref{potshift}) is used. From (\ref{nemui2}), the optical
conductivity in the continuum limit is obtained as
\begin{eqnarray}
\sigma_{\rm res}(\omega) &=& e^2 a^2 {\rm Im}\,
\frac{\widetilde{\Lambda}
({\rm i}\omega_m \rightarrow\omega+{\rm i}\delta)}{\omega} \\
&=& \frac{e^2}{\Mbar} {\rm Re} \, \left.
\frac{1}{z + \hat{\gamma}(z)} \right|_{z\rightarrow -{\rm i}\omega
+\delta} .
\end{eqnarray}
This expression corresponds to the optical conductivity
in the continuum limit, (\ref{contres}).

\section{Ohmic Damping} \label{Ohmic}
\subsection{General review of ohmic damping}
In this section, we consider the ohmic damping case, 
in which the spectral density is described as
\begin{equation}
J(\omega)=\frac{2\pi K}{a^2}\omega {\rm e}^{-\omega/\omega_c}.
\end{equation}
Here, $K$ is a dimensionless coupling constant. 
The Fourier component of the potential 
in (\ref{phidef}) is calculated as
\begin{equation}
\phi({\rm i}\omega_m) = \left\{
\begin{array}{ll} \dis{-\frac{2\pi K}{|\omega_m|}},
& (0<|\omega_m| \ll \omega_c), \\
2K\log \dis{\frac{\beta\omega_c}{2\pi}}, & (\omega_m = 0).
\end{array} \right.
\label{sususu}
\end{equation}
The calculation of $\phi_0 = \phi({\rm i}\omega_m =0)$ 
is given in Appendix~\ref{app1}.
From (\ref{sususu}), the renormalized hopping
amplitude $\Dbar$ in (\ref{Dbardef}) is given by
\begin{equation}
\Dbar = \Delta \left(\frac{\beta\omega_c}{2\pi}\right)^{-K}.
\end{equation} 
In the following subsections, we calculate the specific heat
and the optical conductivity for the ohmic dissipation.
Before showing the results, we briefly review previous 
works on the properties of the ohmic dissipation.

\begin{figure}[tb]
\hfil
\epsfile{file=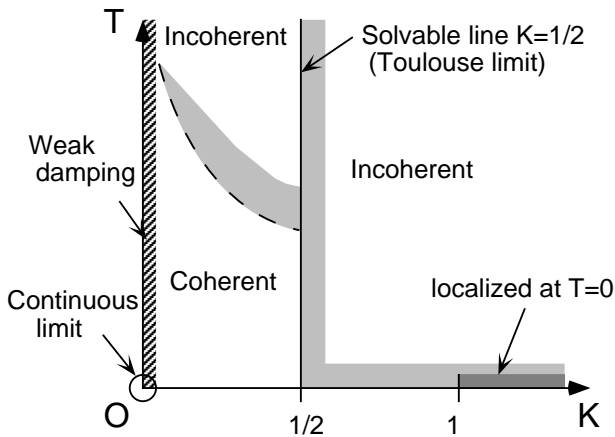,scale=0.7}
\hfil
\caption{The phase diagram of dissipative two-state systems
for the ohmic damping. This phase diagram has been obtained by
the Non-Interacting Blip Approximation adopted 
in ref.~\citen{Leggett87}. In the `coherent' region, 
the expectation value of the position, $\langle q(t) \rangle$ shows
oscillations, while in the `incoherent' region, $\langle
q(t) \rangle$ shows the exponential decay.
The regions which allow analytical treatments is also shown:
the weak coupling regime and the solvable line ($K=1/2$). 
The localization-delocalization transition at $T=0$ is also drawn.}
\label{phase}
\end{figure}

For two-state systems, nonequilibrium behaviors of the
dissipative particle 
have been studied by Leggett~{\it et. al}.~\cite{Leggett87}
They have calculated the real-time evolution of 
the expectation of the particle position with the Non-Interacting
Blip Approximation (NIBA). 
The properties of the system are determined by the damping strength
$K$ and the temperature $T$, where the phase diagram obtained by NIBA
is given in Fig.~\ref{phase}. 
There exist two regions in the phase diagram.
In the `incoherent' region, the particle moves incoherently,
and the expectation value of the position, $\langle q(t)\rangle$
decays exponentially to the thermal equilibrium state.
On the other hand, in the `coherent' region, $\langle q(t)\rangle$
shows damped oscillation. We should note that
several different
definitions of the terms, `coherent' and `incoherent', 
are possible.~\cite{Egger97}
For example, in the numerical study of
equilibrium properties of dissipative two-state 
systems,~\cite{Chakravarty95,Costi96} the term `incoherence' is 
defined by disappearance of an inelastic peak of
the response function, which is observed at $K>0.33$.
A different definition of the term, 
`incoherence' is also used in \S~\ref{OhmicGreen}.

As compared with the two-state systems, it is more difficult to
solve the lattice model studied in this paper.
For example, the simple approximation 
such as NIBA is not applicable to the lattice model.
The dissipative particle in the tight-binding model has been
studied within the real-time path integral formulation
in detail.~\cite{Weiss85,Weiss88,Sassetti92}
Particularly, early discussions are based on the 
the duality transformation, which relates the tight-binding model
to the periodic-potential system in a weak corrugation 
limit.~\cite{Fisher85,Zwerger87,Eckern87,Schmid83}
In these works, only the mobility of the dissipative particle
has been of interest, and the whole dynamical properties of
equilibrium states such as the optical conductivity have not
been considered. Further, the mobility 
at intermediate temperatures has not been obtained exactly 
except for the special value of the damping strength ($K=1/2$).

In this paper, we consider mainly two limiting regions in the phase 
diagram. The first one is the weak damping region $K \ll 1$,
which allows analytical treatment at all temperatures.
The formulation of the weak damping theory can connect
the low-temperature `coherent' region to
the high-temperature `incoherent' region in a natural way.
The result obtained in this narrow region will also be useful to
understand properties in the other region.
The second region is the high-temperature and/or strong 
damping region. In this limit, the dissipative particle moves 
incoherently with the hopping rate determined mainly by the Fermi's
golden rule. Especially, in the region $K>1$, the high temperature
expansion is convergent, and the Fermi's golden rule gives the
leading term at all temperatures. 
On the other hand, for $K<1$, the high temperature
expansion fails below the Kondo temperature $T_{\rm K}$.
The summary of the results on these two region
is given in Table~\ref{resulttable}.

\begin{table}[tb]
\begin{center}
\begin{tabular}{@{\hspace{\tabcolsep}\extracolsep{\fill}}ccc}
\hline
& Low-temperature and 
& High-temperature \\
& weak-damping region 
& or/and strong damping \\
\hline
Specific heat $C$ 
& $\propto T/K$ 
& $\propto \left\{ \begin{array}{ll} T^{-2+2K} & (K<3/2) \\
T & (K>3/2) \end{array} \right.$ 
\\
\hline
Optical conductivity $\sigma(\omega)$ 
& $\propto \left\{
\begin{array}{ll} \omega^{-2+2K} & (\omega \gg \gamma_0) \\
{\rm const.} & (\omega \ll \gamma_0) \end{array} \right.$ 
& $\propto \left\{
\begin{array}{ll} \omega^{-2+2K} & (\omega \gg KT) \\
T^{-2+2K} & (\omega \ll KT) \end{array} \right.$
\\
\hline
DC conductivity $\sigma_{\rm DC}$ 
& $\propto 1 - {\rm const.}T^2$
& $\propto T^{-2+2K}$ \\
\hline
\end{tabular}
\end{center}
\caption{Summary of the results.
Here, $K$ is the dimensionless damping coefficient,
and $\gamma_0$ is the damping frequency defined by (\ref{gamma0def}).
The specific heat at high temperatures is obtained 
from the calculation in two-state systems in ref.~\citen{Goerlich88}.}
\label{resulttable}
\end{table}

With regard to the optical conductivity, 
the result on the continuum limit
is also useful to understand the general behavior.
The continuum limit corresponds to 
the low-temperature and weak-coupling limit ($K, T \rightarrow 0$), and
the particle is expected to move coherently. 
From (\ref{contopteq}), the optical conductivity in
the continuum limit is calculated as a simple Drude form
\begin{equation}
\sigma_{\rm res}(\omega)=
\frac{e^2 \gamma}{\Mbar(\omega^2+\gamma^2)}.
\label{Drudeform}
\end{equation}
When the finite damping is introduced,
the optical conductivity deviates from
the Drude form (\ref{Drudeform}). This behavior
is studied in detail in \S~\ref{weakopt}.

In addition to the above regions, it is expected that
the calculation at $K=1/2$ is tractable analytically by following 
ref.~\citen{Sassetti92}, though we do not deal with the case
in this paper. In the other regions (particularly for $0<K<1$), 
the analytical method cannot be used.
Hence, the dynamical properties of these region
will have to be studied by numerical calculation,
which has not been performed before to the lattice model 
to our knowledge.

Finally, we comment on the renormalization of $\Delta$ due to
the damping. For the ohmic damping, the relevant frequency of 
the system is given by
\begin{equation}
\Deff = \Delta \left(\frac{\Delta}{\omega_c}\right)^{K/(1-K)},
\label{Deffd2}
\end{equation}
for $0<K<1$. The frequency gives the scale of the Kondo temperature
$T_{\rm K}$. 
By rewriting $\Deff$ as $\Delta$, the final results can be expressed
without the cutoff frequency $\omega_c$.~\cite{Leggett87} 
In this paper, all the results are expressed by $\Deff$ finally.

\subsection{Specific heat}\label{heat}
In this subsection, we consider the specific heat of the system
coupled to the ohmic heat bath. 
At high temperatures, we can obtain the specific heat 
$C$ from (\ref{Zhigh}). This result is exactly twice as large as 
the one obtained in dissipative two-state systems.
As a result, the specific heat is proportional to $T^{2K-2}$
for $0<K<3/2$, and proportional to $T$ for $K>3/2$. 
For details of the calculation and result, see
ref.~\citen{Goerlich88}.

\begin{figure}[tb]
\hfil
\epsfile{file=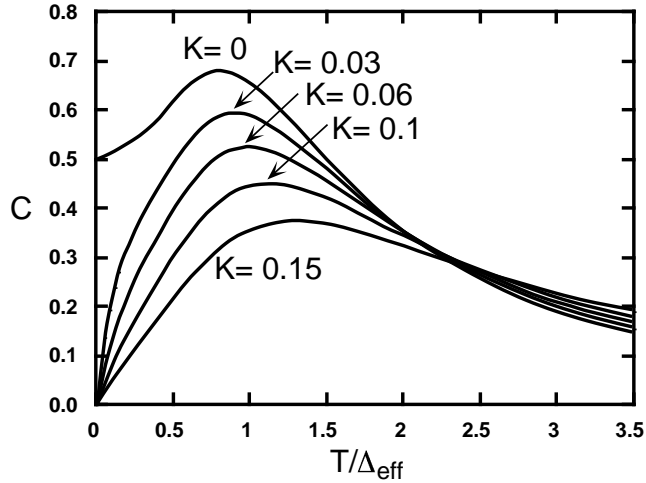,scale=0.65}
\hfil
\caption{The specific heat for weak damping
($K=0.03, 0.06, 0.1, 0.15$). The specific heat is suppressed
to zero as the temperature decreases. This behavior is
quite different from the dissipationless case ($K=0$), where the
specific heat approaches a nonzero constant at low temperatures. }
\label{heatgraph}
\end{figure}
Next, we focus on the weak damping region $K\ll 1$.
From eqs.(\ref{forms})-(\ref{ndef}), we obtain
\begin{eqnarray}
& & Z = \int_0^{2\pi} \frac{{\rm d}\theta}{2\pi} {\rm e}^{U(n)}, \\
& & U(n) = n + \log\Gamma(Kn+1) - Kn\psi(Kn+1), 
\label{Unexp} \\
& & Q(n) = K(\psi(Kn+1)+\gbar),
\label{Qnexp} \\
& & 2\beta\Dbar\cos\theta = n {\rm e}^{-Q(n)},
\label{nthetaexp}
\end{eqnarray}
where $\gbar$ is the Euler's constant, and $\psi(z)$
is the polygamma function.
From this expression, the specific heat is obtained as
Fig.~\ref{heatgraph} by numerical integration.
We can see that, even for the weak damping, the specific heat 
is suppressed at low temperatures, and goes to 
zero in the limit $T\rightarrow 0$. 
This behavior is different from the dissipationless case $K=0$,
where $C$ approaches a nonzero constant value at low temperatures.
This result can be interpreted as follows.
In the dissipationless case, the partition function is
classical in the momentum in the sense that the Hamiltonian is 
diagonalized in the momentum space. 
We note that the quantization of the momentum 
due to a boundary condition is not taken in
this formalism so that the specific heat approaches a classical
nonzero value. When damping is introduced,
the Hamiltonian cannot be diagonalized in the momentum space,
and the quantum effects suppress the specific heat
at low temperatures.

The asymptotic expansion of $Z$ for $\beta =1/T(\gg 1)$ gives
\begin{equation}
\ln Z = 2(1-K)\beta c \Deff + \frac12 \ln K 
+ \frac{1}{24K \beta c\Deff} + {\cal O}(\beta^{-2}),
\label{Zasymp}
\end{equation}
where $c$ is a renormalization factor defined by 
$c=(4\pi K {\rm e}^{\gbar})^{K/(1-K)}$, and takes $c\sim 1$
for weak damping, $K \ll 1$. 
Details of the derivation is given in Appendix~\ref{app3}.
From the partition function, we obtain the low-temperature 
behavior of the system energy $E$ and the specific heat $C$ as
\begin{eqnarray}
E &=& -\frac{\partial}{\partial \beta}
\left( \ln Z \right) = -2(1-K)c \Deff + \frac{1}{24 K \beta^2 c \Deff} 
+ {\cal O}(\beta^{-3}) \\ 
C &=& \frac{\partial E}{\partial T} =
\frac{T}{12K c \Deff} +{\cal O}(T^2),
\label{lowTheat}
\end{eqnarray}
We can see that the specific heat is 
proportional to the temperature $T$, and the coefficient
depends on the damping strength as $C\simeq T/K\Delta_{\rm eff}$.
We denote the relevant energy scale with $\gamma = 4\pi K c \Deff$.
As shown later, $\gamma$ expresses the relevant energy scale 
in the optical conductivity for weak coupling region. 
Then, we can write the specific heat as 
\begin{equation}
C\simeq \pi T/3\gamma
\end{equation}
at low temperatures.

\begin{figure}[tb]
\hfil
\epsfile{file=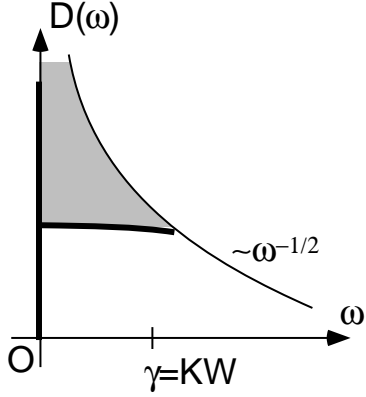,scale=0.75}
\hfil
\caption{The density of states in the dissipative case
(thick line) and the free case (thin line).
By introducing dissipation, it seems that 
the states expressed by the patched area 
moves to the weight of the delta function at $\omega=0$.}
\label{DOSfig}
\end{figure}

Next, we discuss the density of states, $D(\omega)$ 
of the dissipative particle. From the partition function,
$D(\omega)$ is calculated by
\begin{equation}
Z(\beta) = \int_{-\delta}^{\infty} {\rm d}\omega \, 
D(\omega) {\rm e}^{-\beta\omega}.
\label{lap}
\end{equation}
Here, the energy is shifted in order that the ground state energy 
becomes zero, and the infinitesimal quantity $\delta$ is
introduced for convenience. In the dissipationless case,
the density of states of 1D lattice systems is calculated as
\begin{equation}
D(\omega) =\frac{1}{\pi\sqrt{
\dis{\left(\frac{W}{2}\right)^2-\left(\frac{W}{2}
-\omega\right)^2}}},
\end{equation}
where $W$ is a band width.
Particularly, for low-energy states, $D(\omega)$ behaves as
\begin{equation}
D(\omega) \simeq \frac{1}{\pi W}\left(\frac{\omega}{W}
\right)^{-1/2}.
\label{freeDOS}
\end{equation}
In the presence of dissipation, the density of low-energy states 
is strongly modified. To see this, we rewrite the partition function
(\ref{Zasymp}) by using the effective band width $W=2c \Deff$ as
\begin{eqnarray}
Z &\simeq& \sqrt{K} {\rm e}^{1/12K W} 
\nonumber \\
&\simeq& \sqrt{K} \left(
1+ \frac{1}{12K W} + {\cal O}(\beta^{-2}) \right).
\end{eqnarray}
Then, the low-energy form of $D(\omega)$ is obtained 
from (\ref{lap}) as
\begin{equation}
D(\omega)\simeq
\sqrt{K} \delta(\omega) 
+\frac{1}{12 K^{1/2} W} + {\cal O}(\omega) ,
\label{DOSDOS}
\end{equation}
The difference between (\ref{freeDOS}) and (\ref{DOSDOS})
is schematically shown in Fig.~\ref{DOSfig}.
We can see that the dissipative environment
strongly modifies the low-energy states.
Roughly speaking, because of dissipation,
some parts of low-energy states up to $\omega\sim KW$
(the patched region in Fig.~\ref{DOSfig}) 
gathers to $\omega=0$ in the form of the delta function.
Although we cannot discuss the details of this change until
the one-particle Green's function is obtained, it can be conjectured
that the dispersion is modified to be flat by dissipation, and
the wave function of the low-energy states are localized in space.

We note that similar behaviors have
been shown in two-state systems, where the Shottkey form of the
specific heat is modified to the $T$-linear behavior at the 
low temperatures.~\cite{Goerlich88} Also in this case, 
the $T$-linear behavior of the specific heat is induced by
the modification of density of states caused by the environment.

\subsection{Optical conductivity}\label{weakopt}
\subsubsection{The limit of high temperatures and/or
strong dissipation}
\begin{figure}[tb]
\hfil
\epsfile{file=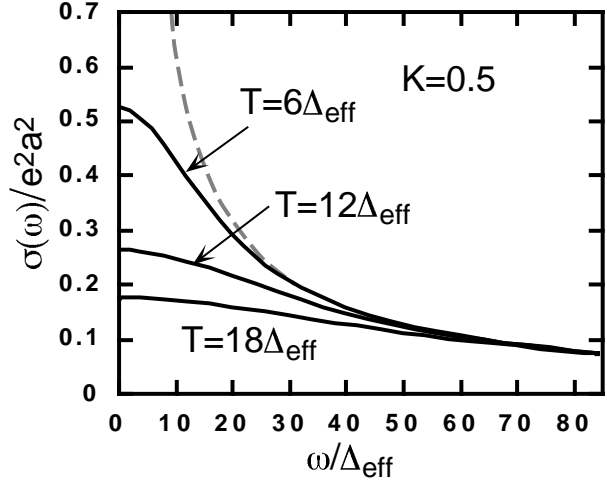,scale=0.75}
\hfil
\caption{The optical conductivity 
$\sigma_{\rm res}(\omega)$  at
high temperatures ($T=6\Deff, 12\Deff, 18\Deff$) for $K=0.5$.
The dashed gray line represents the asymptotic behavior 
($\propto 1/\omega$) at $\omega/
\Deff\gg 1$. The characteristic energy at which the optical
conductivity deviates from the asymptotic form is estimated
as $KT$.}
\label{opt}
\end{figure}

For the ohmic damping, $S(t)$ and $R(t)$ in 
(\ref{Stdef})-(\ref{Rtdef}) are obtained as
\begin{eqnarray}
S(t) &=& 2K\log\left|\frac{\beta \omega_c}{\pi}\sinh\left(
\frac{\pi t}{\beta} \right) \right|, \\
R(t) &=& \pi K {\rm sign}(t).
\end{eqnarray}
From (\ref{highcorr2}), we obtain the optical conductivity 
$\sigma_{\rm res}(\omega) = e^2 a^2 {\rm Im} \Lambda(\omega)
/\omega$ as
\begin{eqnarray}
\sigma_{\rm res} (\omega) 
&=& 2e^2 a^2
\frac{\Delta^2}{\omega_c}\left(\frac{\beta\omega_c}{2\pi}
\right)^{1-2K} \frac{|\Gamma(K+{\rm i}\beta\omega/2\pi)|^2}
{\Gamma(2K)} \frac{\sinh\beta\omega/2}{\omega} \\ 
&=& 2 e^2 a^2\Deff \left(\frac{\beta \Deff}
{2\pi}\right)^{1-2K} \frac{|\Gamma(K+{\rm i}\beta \omega / 2\pi)|^2}
{\Gamma(2K)} \frac{\sinh\beta\omega/2}{\omega}.
\end{eqnarray}
The temperature and frequency dependences of $\sigma_{\rm res}(\omega)$ 
are shown in Fig.~\ref{opt}.
In the low-frequency side ($\omega \ll K T$),
$\sigma_{\rm res}(\omega)$ depends on the temperatures.
The DC conductivity 
$\sigma_{DC} = \sigma_{\rm res}(\omega\rightarrow 0)$
is calculated as
\begin{equation}
\sigma_{DC} = 2\pi e^2 a^2
\left(\frac{\beta\Deff}{2\pi}\right)^{2-2K}
\frac{\Gamma(K)^2}{\Gamma(2K)}, 
\label{highTDC}
\end{equation}
which is proportional to $T^{2K-2}$.
On the other hand, in the high-frequency limit, 
$\sigma_{\rm res}(\omega)$ takes a temperature-independent form
\begin{equation}
\sigma_{\rm res}(\omega)
=\frac{2\pi e^2 a^2}{\Gamma(2K)} \left(
\frac{\Delta_{\rm eff}}{\omega} \right)^{2-2K}.
\label{highThighw}
\end{equation}
On the high-frequency side, $\sigma_{\rm res}(\omega)$ is proportional
to $\omega^{-2+2K}$, and deviates from the Drude
form (\ref{Drudeform}).

\subsubsection{Weak coupling region}
Next, we consider the weak damping region, $K\ll 1$.
For the ohmic damping, the screened potential
defined by (\ref{screenpotdef}) is calculated as
\begin{equation}
\vphi(\tau;\theta) = \frac{1}{\beta} \sum_{\omega_m} \frac{-2\pi K}
{|\omega_m|+\gamma(\theta)} {\rm e}^{{\rm i}\omega_m \tau},
\label{optsum}
\end{equation} 
where $\gamma(\theta)$ is an inverse of a screening length 
in the $\tau$-direction given by
\begin{equation}
\gamma(\theta) = 2\pi K n(\theta)/\beta.
\label{gammadefz}
\end{equation}
Following the usual way, 
the sum over the Matsubara frequency
in (\ref{optsum}) is replaced by an integral form.
As a result, $\vphi({\rm i}t;\theta) = S(t;\theta) + {\rm i}
R(t;\theta)$ is obtained as
\begin{eqnarray}
R(t;\theta) 
&=& \pi K {\rm e}^{-\gamma(\theta) t} 
\label{Sweak}\\
S(t;\theta) &=& 2K \int_0^{\infty} {\rm d}\omega \frac{\omega}{\omega^2
+\gamma(\theta)^2}\left[ \frac{2}{\beta\omega} -\coth\frac{\beta\omega}{2}
\cos\omega t\right] \nonumber \\
&+& 2\pi K\Theta(-\gamma) \left[\frac{2}{\beta\omega} 
-\coth\frac{\beta\omega}{2} \cos\omega t \right], 
\label{Rweak}
\end{eqnarray}
where $\Theta(x)$ is a step function. For no damping case
$\gamma\rightarrow 0$, the above expressions become
unscreened potentials, $S(t)$ and $R(t)$
in (\ref{Stdef})-(\ref{Rtdef}). 
Analytical expressions for $S(t;\theta)$ are obtained only for
two limiting cases. At high temperatures $\beta\rightarrow 0$,
$S(t;\theta)$ is calculated as
\begin{equation}
S(t;\theta) = \frac{2\pi K}{\beta\gamma(\theta)} 
(1-{\rm e}^{-\gamma(\theta) t}).
\end{equation}
On the other hand, at low temperatures, $S(t;\theta)$
is expanded by the temperature $T=1/\beta$ as
\begin{eqnarray}
S(t;\theta) &=& S_0(t;\theta) + \frac{2\pi K T}{\gamma(\theta)} 
+ {\cal O}(T^2), \\
S_0(t;\theta) &=& K \left[ {\rm e}^{\gamma(\theta) t} 
{\rm Ei}(-\gamma(\theta) t)
+ {\rm e}^{-\gamma(\theta) t} {\rm Ei}(\gamma(\theta) t) \right],
\end{eqnarray}
where ${\rm Ei}(z)$ is the exponential integral function, and
$S_0(t)$ is a temperature-independent part.
In both cases, the limiting value at $t\rightarrow \infty$ 
is given by
\begin{equation}
S(t\rightarrow\infty; \theta) = \frac{2\pi K T}{\gamma(\theta)}.
\label{Stinf2}
\end{equation}

\begin{figure}[tb]
\hfil
\epsfile{file=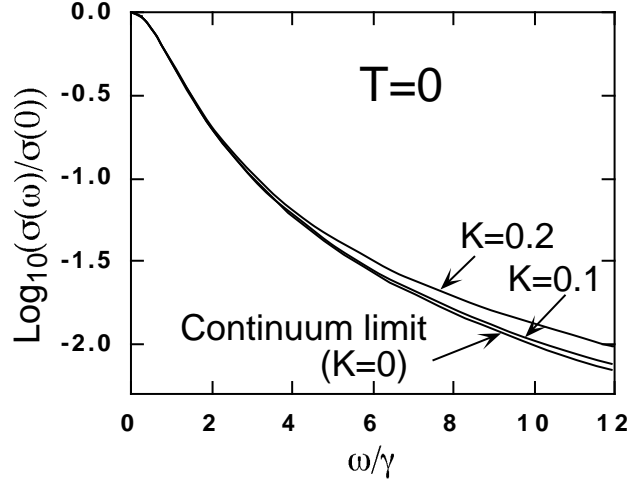,scale=0.75}
\hfil
\caption{The optical conductivity $\sigma(\omega)$
at zero temperature for weak damping ($K=0, 0.1, 0.2$).
As the damping increases, $\sigma(\omega)$ deviates from 
the Drude form ($K=0$). At high frequency, $\sigma(\omega)$
is proportional to $\omega^{2K-2}$.}
\label{zeroopt}
\end{figure}

Although we can calculate $\sigma(\omega)$ 
by the expressions (\ref{Sweak})-(\ref{Rweak})
for all temperatures, we only consider zero temperature
in this paper. At zero temperature, 
the dominant contribution comes from $\theta=0$
in the integral (\ref{weakformalism}). 
The real-time correlation function for $K\ll 1$ is obtained 
by using the asymptotic form $n=2\beta c \Deff$ as
\begin{equation}
\Lambda(\omega) = \frac{\gamma_0}{2\pi K} \int_0^{\infty}
{\rm d}t \, {\rm e}^{{\rm i}\omega t - \gamma_0 t}
\cosh\left[ S_0(t)\right].
\label{zeroopt2}
\end{equation}
Here, 
$\gamma_0=\gamma(\theta=0)$ is the damping frequency at $T=0$ given by
\begin{equation}
\gamma_0 = 4\pi K c \Deff,
\label{gamma0def}
\end{equation}
where $c=(4\pi K {\rm e}^{\gbar})^{K/(1-K)}$.
The optical conductivity can be calculated
by numerical Fourier transformation of (\ref{zeroopt2}).
The result for $\sigma_{\rm res}(\omega)$ is shown in 
Fig.~\ref{zeroopt}. The Drude formula (\ref{Drudeform}) in
the continuum limit is also shown. At the high frequency side,
the optical conductivity of the dissipative system deviates from
the Drude form.

The qualitative behavior of $\sigma(\omega)$ is
obtained by considering the limiting from of $S_0(t)$:
\begin{equation}
S_0(t) = \left\{ \begin{array}{ll}
2K \log ({\rm e}^{\gbar}\gamma_0 t) & (t \rightarrow 0), \\
\dis{\frac{2K}{(\gamma_0 t)^2}} & (t \rightarrow \infty), 
\end{array} \right.
\label{ffff}
\end{equation}
where $\gbar$ is the Euler's constant.
From (\ref{zeroopt2}) and (\ref{ffff}), 
the limiting behavior of the optical conductivity
is obtained as
\begin{equation}
\sigma(\omega) = \left\{ \begin{array}{ll}
\dis{\frac{2\pi e^2 a^2}{\Gamma(2K)} \left( \frac{\Deff}{\omega} 
\right)^{2-2K}},
& (\omega \gg \omega^*), \\
\dis{\frac{e^2 a^2}{2\pi K} \frac{\gamma_0^2}{\gamma_0^2+\omega^2}},
& (\omega \ll \omega^*),
\end{array} \right.
\label{Tosigmaeq}
\end{equation}
where $\omega^*=(2c)^{1/2K}$ is the crossover frequency at which
the two expressions take the same value.
The high-frequency behavior of $\sigma(\omega)$
agrees with the result of the high temperature limit (\ref{highThighw}).
This indicates that the high-frequency side is
independent of the temperature at {\it any} damping strength,
since (\ref{highThighw}) is valid at any value of $K$.
On the other hand, the low-frequency side coincide with the Drude form
(\ref{Drudeform}) obtained in the continuum limit.
It is interesting that 
the asymptotic form in the limit $\omega\rightarrow\infty$ 
of the Drude form (the first equation in (\ref{Tosigmaeq}) )
deviates from the high frequency form (the second equation
in (\ref{Tosigmaeq}) ) shown
in the lower equation of (\ref{Tosigmaeq}) by factor 2.
At first glance, one might have suspicion on this difference.
However, this difference indeed exists: When we take the limit
$K\rightarrow 0$, the crossover frequency $\omega^*$ in 
(\ref{Tosigmaeq}) goes to infinity. Then, the region
$\omega\gg\omega^*$ disappears and the lower expression
in (\ref{Tosigmaeq}) governs even the high-frequency region,
which reproduces (\ref{Drudeform}).

On the high frequency side $\omega \gg \omega^*$, 
the particle moves by using the heat-bath excitations.
Therefore, the particle moves incoherently
even at zero temperature. The results are relevant to
the photo-assisted tunneling.~\cite{Chakravarty85}

\subsubsection{DC conductivity}
We consider the DC conductivity $\sigma_{\rm DC}$
of the lattice model for the weak coupling region ($K \ll 1$).
The low-temperature behavior
is obtained by the asymptotic expansion over $\beta=1/T$ as
\begin{equation}
\sigma_{\rm DC}/\sigma_{\rm DC}^0 = 
1-\frac18 \left(\frac{T}{c \Deff}\right)^2
+{\cal O}(T^3),  
\label{dccondT}
\end{equation}
where $\sigma_{\rm DC}^0 = e^2 a^2/2\pi K$.
(See Appendix~\ref{app3}.)
Thus, the thermal fluctuation suppresses $\sigma_{\rm DC}$.
The $T^2$-suppression at low temperatures
is consistent to the general study of
the linear mobility $\mu_l =\sigma_{\rm DC}/e^2$.~\cite{Weiss91}
In the previous work, the linear mobility has not been obtained 
correctly even in the weak damping $K\ll 1$.~\cite{comment}
The high-temperature limit of $\sigma_{\rm DC}$ for $K\ll 1$
is obtained from (\ref{highTDC}) as
\begin{equation}
\sigma_{\rm DC}/\sigma_{\rm DC}^0 =
8 \pi^2 \left( \frac{\beta\Deff}{2\pi} 
\right)^{2-2K}.
\end{equation}

We compare this result with the linear mobility
$\mu_l$ obtained in ref.~\citen{Weiss91}. For the special value
$K=1/2$, the linear mobility is calculated for all
temperatures as
\begin{equation}
\mu_l = \frac{\mu_0 \Deff}{2T} 
\psi'(\Deff/2T +1/2),
\end{equation}
where $\psi'(z)$ is the trigamma function,
and $\mu_0=a^2/2\pi K$. At low temperatures,
the DC conductivity $\sigma_{\rm DC}=\mu_l/e^2$ 
behaves as
\begin{equation}
\sigma_{\rm DC} / \sigma_{\rm DC}^0
\simeq 1 - \frac{T^2}{3\Deff^2},
\end{equation}
and show the $T^2$-suppression.
This strongly indicates 
that the $T^2$-suppression appears
for any value of $0<K \le 1/2$. 
At high temperatures, the DC conductivity for $K=1/2$ shows
a $1/T$ behavior as
\begin{equation}
\sigma_{\rm DC}/\sigma_{\rm DC}^0 \simeq \frac{\pi^2 \Deff}{4T},
\end{equation}
which agrees with (\ref{highTDC}) for $K=1/2$.

\subsection{Application of the results to real systems}
\label{OhmicGreen}

The tight-binding model of a particle with ohmic damping 
is relevant to following systems: the muon motion in a metal, 
the incoherent carrier motion in coupled Tomonaga-Luttinger
liquid and Josephson junction systems.
We, however, must comment on the qualification in applying 
the result obtained in this section to each system.

The muon motion in a metal has been discussed first in connection
with the Anderson's orthogonality theorem,~\cite{Kondo76,Kondo84,
Yamada84} and subsequently this problem has been formulated
based on the path integral.~\cite{Sols87} 
Experiments in Cu have showed the $T^{2K-1}$-behavior in the
diffusion rate as expected in the incoherent regime
of the Caldeira-Leggett model. In this experiment,
the effective hopping amplitude 
$\Delta_{\rm eff}\sim 0.1{\rm mK}$ is
so small that the low-temperature
region is difficult to be observed in the presence of 
disorder.~\cite{Kondo88} 
Moreover, the Caldeira-Leggett (CL) model is thought to fail
in describing the heavy particle motion at low temperatures,
since the damping is overestimated
as pointed out by several authors.~\cite{Sols87,Zimanyi87} 
The authors have showed that 
the damping strength due to the conduction electrons
becomes small at long distances, while
in the CL model, this feature is absent, and the damping is
overestimated at long distances.
This difference would be critical particularly at low temperatures,
since the configuration paths with large spatial 
distribution are significant.

Recently, the CL model has been used to 
discuss the incoherent non-metallic transport 
in strongly correlated anisotropic metal.~\cite{Clarke97} 
Based on two coupled Tomonaga-Luttinger(TL) liquids model,
the authors have discussed interliquid particle motion
with small hopping matrix element, by taking a dissipative
two-level system as an effective model.
It is, however, problematic to extend this treatment 
if infinite number of TL-liquids are coupled.
The effective Hamiltonian does not correspond to
the Caldeira-Leggett model, because the damping coefficient
again should depend on distance between chains.
Hence, it seems  
also problematic to take the naive CL model as an effective model
of coupled TL-liquids for calculating low-energy properties
at low temperatures, and the application of the tight-binding model
for incoherent interchain hopping should be considered more carefully.

The dissipative phase motion in Josephson 
junctions~\cite{Zwerger87,Schoen90,Kato96} seem to be
free from the problems in contrast to the previous two systems,
and the result obtained in this section may be applicable. 
In this system, however, damping is so strong ($K\sim 1$)
that the coherent motion has not been confirmed yet. 
In this system,
the incoherent tunneling motion with the rate proportional 
to $T^{2K-1}$ has been observed,~\cite{Han91} while
macroscopic quantum coherence~\cite{Leggett87} 
in the two-state systems expected 
in the weak-damping region ($K\ll 1$) 
has not been observed so far.

Thus, unfortunately, no system is at the moment available
to directly apply the result in this paper.
We wish that progresses in experimental techniques 
would make it possible to realize experiments
to which the result in this paper can be applied.

We also speculate
that the present model may be applicable to the incoherent 
carrier motion in strongly correlated electron systems.
For example, the retracable path approximation for
correlated electron systems~\cite{Brinkman78,Rice89,Metzner92}
gives the optical conductivity 
of non-Drude forms roughly proportional to $1/\omega$
in contrast to the Drude form $\sigma(\omega)\sim 1/(\omega^2+
\gamma^2)$. This behavior can be reproduced in the Caldeira-Leggett
model for the ohmic damping case ($s=1, K\sim1/2$) or
the superohmic damping case ($s\sim2$).
In the retracable approximation, 
the coherent motion of the particle is strongly
suppressed by the antiferromagnetic background. This mechanism is
similar to the present model of the dissipative particle,
when the heat bath is regarded as the antiferromagnetic background.
The similarity is, however, not complete, since the density of states,
$D(\omega)$ for the ohmic case given in (\ref{DOSDOS}) 
differs from the result obtained by the retractable path
approximation at low-energy states. 
Nevertheless, we speculate that the lattice model
of the dissipative particle may be utilized as a basic
model of the decoherence of the particle due to the 
electron-electron correlation at least in the high energy
region of the optical conductivity.

\section{Summary}\label{Summary}
We have studied the thermodynamics and transport properties 
of the dissipative particle in the tight-binding model.
By utilizing the imaginary-time path integral,
the specific heat and the optical conductivity have been formulated 
for arbitrary form of the spectral density.
A systematic approximation has been considered to treat
the weak-coupling region at all temperatures.
We have obtained an analytical form which can connect
the high-temperature region to the low-temperature region.

The actual calculation
has been performed for the ohmic damping case. 
The specific heat at weak damping 
shows $T$-linear behaviors at low temperatures, and
the density of low-energy states is modified by dissipation.
The optical conductivity shows
non-Drude from even at zero temperature, and
behaves as $\omega^{2K-2}$ on the high-frequency side,
where $K$ is the dimensionless damping strength.
At high temperatures, the high-frequency side does not change the
non-Drude form, and only the low-frequency side depends 
on the temperature $T$.
Particularly, the DC conductivity is proportional to $T^{2K-2}$. 
We also comment on the application of the Caldeira-Leggett theory
to real systems including the 
descriptions of the incoherent transport in 
correlated electron systems.

In this paper, we have focused only on the ohmic damping.
The detailed calculation for the non-ohmic damping and the 
solvable line $K=1/2$ remains for future study. 
We expect that the simple model considered in this paper
may contribute to understand transport properties
of a particle coupled to other degrees of freedom,
and wish that the results of this paper
may be directly observed in a properly fabricated experiment.
   \appendix
\section{Non-ohmic Damping in the Incoherent Tunneling Regime}
\label{app0}
In this appendix, we study the optical conductivity 
$\sigma(\omega)$ for
the non-ohmic damping case in the incoherent regime
following the formulation in \S~\ref{fhigh}.
The results on the tunneling rate $\Gamma(\eps)$ 
obtained in ref.~\citen{Leggett87} are utilized 
to obtain $\sigma(\omega)$. We only consider a few 
limiting cases which allow analytical treatment.

We begin with the superohmic damping $1<s<2$. 
Under the strong bias $\eps\gg\Dbar$ and
the weak coupling condition $\delta_s
(\eps/\widetilde{\omega})^{s-1}\ll 1$,
the tunneling rate $\Gamma(\eps)$ is calculated as
\begin{equation}
\Gamma(\eps) = \frac{a^2}{2}
\left(\frac{2\Dbar}{\eps}\right)^2 J(\eps) \coth(\beta\omega/2).
\end{equation}
By using (\ref{Jomegad}) and
(\ref{incoptform}), we obtain the optical conductivity as
\begin{equation}
\sigma(\omega) = 4\pi \delta_s \left(\frac{\Dbar}{\omega}\right)^2
\left(\frac{\omega}{\widetilde{\omega}}\right)^{s-1},
\label{superhighw}
\end{equation}
which is temperature-independent at all temperatures.
Hence, the optical conductivity behaves as $\sigma(\omega)\propto
\omega^{s-3}$. This behavior corresponds to the result
(\ref{contqb}) in the continuum limit ($T,\delta_s \rightarrow 0$)
on the high frequency side, except a difference by a factor 2.
Similar difference appears also in the case of the ohmic damping,
and the origin of the factor 2 is expected to be the same.
(See (\ref{Tosigmaeq}) and the following text.)

The tunneling rate for $\eps=0$ is calculated as follows:
\begin{equation}
\Gamma = \left( \frac{(2-s)\sin(\pi s/2)}{2\delta_s \Gamma
(s-1) \sin \pi(s-1)} \right)^{1/(2-s)}
\Gamma\left(\frac{3-s}{2-s}\right) \frac{\Dbar^2}
{\widetilde{\omega}} \left(\frac{\widetilde{\omega}}{T} 
\right)^{1/(2-s)}.
\label{superDC}
\end{equation}
This result is valid at high temperatures $T^*\ll T(\ll \omega_c)$,
where $T^*$ is given by
\begin{equation}
T^* = \frac{\Dbar}{\delta_s \Gamma(s-1)} \left(
\frac{\widetilde{\omega}}{\Dbar}\right)^{s-1}.
\end{equation}
The DC conductivity is obtained from (\ref{superDC})
by $\sigma_{\rm DC}= e^2 a^2 \Gamma / 2T$, and
behaves as $\sigma_{\rm DC}\propto T^{-\frac{3-s}{2-s}}$
at high temperatures. The exponent of $T$ decreases
from $-2$ to $-\infty$ as increasing $s$ from 1 to 2.

Next, we consider the subohmic damping $0<s<1$.
At zero temperature, the tunneling rate is calculated 
under the condition $\delta_s
(\eps/\widetilde{\omega})^{s-1}\gg 1$ as
\begin{eqnarray}
\Gamma(\eps) &=& \frac{(B\Delta)^2}{4\widetilde{\omega}}
\left(\frac{2\pi[2\delta_s\Gamma(s)]^{1/2}}{s}\right)^{1/2}
\left(\frac{\widetilde{\omega}}{\eps}\right)^{(1+s)/2s} 
\nonumber \\
& & \times \exp \left[ -\frac{s}{1-s}[2\delta_s\Gamma(s)]^{1/s}\left(
\frac{\widetilde{\omega}}{\eps} \right)^{(1-s)/s} \right],
\end{eqnarray}
where $B=\exp[ \, \delta_s | \Gamma(s-1) | (\widetilde{\omega}/
\omega_c)^{1-s} \, ]$. This result cannot be related to
the result of the continuum limit because of the 
strong coupling condition $\delta_s
(\eps/\widetilde{\omega})^{s-1}\gg 1$.
The optical conductivity is obtained by $\sigma(\omega)
= a^2 e^2 \Gamma(\omega)/\omega$, and behaves as $\sigma(\omega)
\sim \omega^{-(1+3s)/2s}
\exp(-{\rm const.} \omega^{-(1-s)/s} )$. Notice that 
the optical conductivity vanishes with an essential
singularity as $\eps\rightarrow 0$.

The tunneling rate for $\eps=0$ for the
condition $\delta_s(\widetilde{\omega}/T)^{1-s}\gg 1$ is 
given by
\begin{eqnarray}
\Gamma &=& \frac{(B\Delta)^2}{2\widetilde{\omega}}
\left(\frac{\widetilde{\omega}}{T}\right)^{(1+s)/2}
\left(\frac{\pi}{2(1+s)^2\Gamma(s)\zeta(1+s)\delta_s}\right)^{1/2} 
\nonumber\\
& & \times \exp \left[ -\frac{1+s}{1-s} \Gamma(s) \left(
\frac{1}{2(1+s)\zeta(1+s)}\right)^{(1-s)/(1+s)}\delta_s
\left(\frac{\widetilde{\omega}}{T}\right)^{1-s} \right].
\label{subDC}
\end{eqnarray}
From (\ref{subDC}), the DC conductivity is obtained by
$\sigma_{\rm DC}=a^2 e^2 \Gamma/ 2T$, and behaves as
$\sigma_{\rm DC}\sim T^{-(3+s)/2}\exp(-{\rm const.} 
\times T^{s-1} )$. The DC conductivity vanishes with an
essential singularity as $T\rightarrow 0$, and this
result is consistent with the continuum limit,
where $\sigma_{\rm reg}(\omega\rightarrow 0)= 0$.
Note that the power law $\sigma_{\rm DC}\sim T^{2K-2}$ 
for the ohmic case $s=1$ is obtained
from (\ref{subDC}) by
carefully considering the limit $s\rightarrow 1^{-}$.

\section{Calculation of $\phi_0$}\label{app1}
In this appendix, we evaluate the zero frequency component
$\phi_0=\phi({\rm i}\omega_m=0)$. From (\ref{form4})
and (\ref{phidef})-(\ref{Domega}),
$\phi_0$ is explicitly given as
\begin{equation}
\phi_0 = \frac{a^2}{\pi} \int_0^{\infty}
{\rm d}\omega
\frac{J(\omega)}{\omega^2}\left( \coth\frac{\beta\omega}{2} -
\frac{2}{\beta \omega} \right).
\label{phi0def}
\end{equation}
We consider the analytical form of the spectral density, 
\begin{equation}
J(\omega) = \frac{2\pi \delta_s}{a^2} 
\left(\frac{\omega}{\widetilde{\omega}}
\right)^{s-1} \omega {\rm e}^{-\omega/\omega_c},
\label{Jomega2}
\end{equation}
where the exponential cut-off is adopted, 
and the cutoff frequency $\omega_c$ is taken as
$\Delta, 1/\beta \ll \omega_c$.
By substituting (\ref{Jomega2}) to (\ref{phi0def}), we obtain
\begin{eqnarray}
\phi_0 &=& 2\delta_s \left(\frac{\omega_c}{\widetilde{\omega}}
\right)^z \Gamma(z)\left[ \frac{2}{(\beta\omega_c)^z}
\zeta(z,1/\beta\omega_c)-1-\frac{2}{z-1}\frac{1}{\beta\omega_c}
\right] \label{Zzzz} \\
&\simeq& 2\delta_s \left(\frac{\omega_c}{\widetilde{\omega}}\right)^z
\Gamma(z) \left[ 1+ \frac{2\zeta(z)}{(\beta\omega_c)^z}+{\cal O}
(1/\beta\omega_c)\right],
\label{phi0exp}
\end{eqnarray}
where $z=s-1$, and $\zeta(z,a)$ is the generalized zeta function,
and $\zeta(z)$ is the zeta function. In the second equation, we have
left only the leading terms under the condition $\beta\omega_c \gg 1$.
For superohmic case $z>0$, the first term in the bracket is dominant, 
and $\phi_0$ is independent of temperatures.
Further, the result is expressed generally
reproduced by the Frank-Condon factor calculated as
\begin{equation}
\phi_0 = \frac{a^2}{\pi}\int_0^{\infty}\frac{J(\omega)}{\omega^2}
{\rm d}\omega,
\end{equation}
when the integral is convergent as in the case of the
superohmic damping $s>1$.
On the other hand, for subohmic damping $z=s-1<0$, 
the second term in the bracket in (\ref{phi0exp}) is dominant.
Hence, $\phi_0$ depends on temperatures.
For the ohmic case $s=1$, the expression (\ref{phi0exp}) fails,
and careful treatment of (\ref{Zzzz}) is required. As a result,
the leading term in $\phi_0$
depends logarithmically on temperatures as
\begin{equation}
\phi_0 \simeq 2K\log \frac{\beta\omega_c}{2\pi},
\end{equation}
where $K=\delta_1$.
In this case, the renormalized matrix element 
(\ref{Dbardef}) is given by
\begin{equation}
\Dbar = \Delta \left(\frac{\beta\omega_c}{2\pi}\right)^{-K}.
\end{equation}
At a first glance, one might regard that 
this result is unphysical, because the matrix element
is reduced to zero as the temperature decreases even for 
the weak coupling region $K\ll 1$.
We, however, show that the renormalization factor 
$(\beta\omega_c /2\pi)^{-K}$
is canceled by other factors for observables
in the weak coupling theory. 

\section{Ring Approximation}\label{app2}
In this appendix, we give details of the calculation
in the weak coupling theory (\S~\ref{weak}) for 
the arbitrary form of $J(\omega)$. We first study 
the partition function
based on the cluster expansion of classical imperfect 
gas.~\cite{Mayler40} The partition function and
the optical conductivity 
are formulated by the ring approximation.

\begin{figure}[tb]
\hfil
\epsfile{file=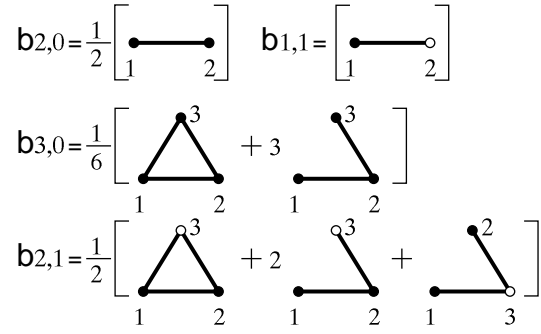,scale=0.5}
\hfil
\caption{The diagrams representing the integrals $b_{l,m}$ 
in (\ref{bstart})-(\ref{bend}). The closed (open) circles
denotes the positive (negative) charges.
Note that $b_{l,m}=b_{m,l}$.}
\label{graph1}
\end{figure}

We begin with the partition function (\ref{clZ}). 
After the potential shift (\ref{potrenorm}), we obtain
\begin{eqnarray}
Z &=& \sum_{m=0}^{\infty} \frac{\Dbar^{2m}}{m!m!}
\prod_{l=1}^{m} \int_0^{\beta} {\rm d}\tau_l
\int_0^{\beta} {\rm d}\rho_l \nonumber \\
&\times& \exp \left[
\sum_{k<l}^{m} \left\{ \phi(\tau_l - \tau_k) 
+ \phi(\rho_l-\rho_k) \right\} - \sum_{l=1}^m \sum_{k=1}^m
\phi(\tau_l-\rho_k) \right\},
\label{clZ2}
\end{eqnarray}
where $\Dbar$ is a renormalized transition amplitude defined
by (\ref{Dbardef}).
Here, new integral variables, $\{\tau_l\}$ and $\{\rho_l\}$
describe the positions of positive ($\xi_l=+1$) and negative 
($\xi_l=-1$) charges. Note that the partition function
is multiplied by the factor $2m!/m!m!$. This factor comes from 
the number of the ways in which $2m$ charges are divided into 
two groups consisting of $m$ charges to guarantee the
electroneutrality condition (\ref{constraint}).
Then, we consider the cluster expansion for the partition 
function (\ref{clZ2}).
To simplify the expansion, we rewrite it 
as the following form
\begin{equation}
Z = \sum_{m=0}^{\infty} \frac{(\beta \Dbar)^{2m}}{m!m!} 
\langle \prod_{k<l}^m (1+f_{kl}^{++}) (1+f_{kl}^{--})
\prod_{k=1}^m \prod_{l=1}^m ( 1+f_{kl}^{+-} ) \rangle.
\label{clZ1}
\end{equation}
Here, the potential $\phi(\tau)$ is replaced by Mayer's functions
\begin{eqnarray}
f_{kl}^{++} &=& {\rm e}^{\phi(\tau_k - \tau_l)}-1, \\
f_{kl}^{--} &=& {\rm e}^{\phi(\rho_k - \rho_l)}-1, \\
f_{kl}^{+-} &=& {\rm e}^{-\phi(\tau_k-\rho_l)}-1, 
\end{eqnarray}
and $\langle \cdots \rangle$ denotes the average
\begin{equation}
\langle \cdots \rangle = \frac{1}{\beta^{2m}} 
\prod_{l=1}^m 
\int_0^{\beta} {\rm d}\tau_l 
\int_0^{\beta} {\rm d}\rho_l \, ( \cdots ).
\end{equation}
Each term in the expansion 
on $f_{kl}$ in (\ref{clZ1}) is expressed 
by a product of integrals of clusters. 
We introduce cluster integrals 
\begin{equation}
b_{l,k} = \frac{1}{l!k!} \sum_{{\rm clusters}} 
\langle \prod_{i,j} f_{ij}^{\sigma_i \sigma_j} \rangle,
\end{equation}
where $\sum_{{\rm clusters}}$ denotes the sum of all
clusters including $k$ positive charges and $l$ negative charges.
The products are taken over all bonds combining $i$-th and
$j$-th charges in each cluster, where $\sigma_i$ and $\sigma_j$
are signs of charges. For example, the explicit form of 
$b_{l,k}$ for small $k$, $l$ is given by
\begin{eqnarray}
b_{0,0} &=& 0, \hspace{5mm} b_{10}=b_{01} = 1, 
\label{bstart} \\
b_{2,0} &=& \frac{1}{2} \langle f_{12}^{++} \rangle, \\
b_{1,1} &=& \langle f_{12}^{+-} \rangle, \\
b_{3,0} &=& \frac{1}{6} \langle f_{12}^{++} f_{23}^{++} 
f_{31}^{++} + 3 f_{12}^{++} f_{23}^{++}  \rangle, \\
b_{2,1} &=& \frac{1}{2} \langle f_{12}^{++} f_{23}^{+-}
f_{13}^{+-} + 2 f_{12}^{++} f_{23}^{+-} + f_{13}^{+-} f_{23}^{+-}
\rangle, 
\label{bend}
\end{eqnarray}
and so on. The above integrals can be viewed by graphical
representations as shown in Fig.~\ref{graph1}.
By the cluster integrals $b_{l,k}$,
the partition function is expressed as
\begin{eqnarray}
Z&=& \sum_{m=0}^{\infty} \frac{(\beta\Dbar)^{2m}}{m!m!}
\sum_{\{m_{l,k}\}'} \frac{m!}
{\dis{\prod_{l,k}m_{l,k}!(l!)^{m_{l,k}}}}
\nonumber
\\ &\times&  \frac{m!}
{\dis{\prod_{l,k}m_{l,k}!(k!)^{m_{l,k}}}}
\times
\prod_{l,k} (l!k!b_{l,k})^{m_{l,k}} \times \prod_{l,k} m_{l,k}!,
\label{weakZ2}
\end{eqnarray}
where $m_{l,k}$ is the number of clusters including $l$ positive
charges and $k$ negative charges. 
The sums and products are taken
over $l$ and $k$ except for the case $(l,k)=(0,0)$.  
The prime in $\{m_{l,k}\}'$ denotes the summation in accordance with
the conditions
\begin{eqnarray}
\sum_{l,k} l m_{l,k} &=& m, \label{const1} \\
\sum_{l,k} k m_{l,k} &=& m. \label{const2}
\end{eqnarray}
The first (second) factor in the sum $\sum_{\{m_{l,k}\}'}$ 
in (\ref{weakZ2}) gives the number of ways
in which $m$ positive (negative)
charges can be distributed into the clusters.
The third part is the relevant 
cluster integrals, and the forth part is the number of
ways to combine $m_{l,k}$ groups of positive charges 
to $m_{l,k}$ groups of negative charges.

From (\ref{const1})-(\ref{const2}), we obtain the constraint
on the total number of charges as
\begin{equation}
\sum_{l,k} (l+k) m_{l,k} = 2m,
\end{equation}
and this constraint can be removed by the sum over $m$.
On the other hand, the electroneutrality condition
\begin{equation}
\sum_{l,k} (l-k) m_{l,k} =0,
\label{consttt}
\end{equation}
remains. To remove the constraint (\ref{consttt}),
we introduce an integral over a new variable $\theta$ as
\begin{equation}
\sum_{m=0}^{\infty} \sum_{\{m_{l,k}\}'} (\cdots)
=  \sum_{\{m_{l,k}\}}  \int_0^{2\pi} 
\frac{{\rm d}\theta}{2\pi} \exp\left({\rm i}\sum_{l,k} (l-k)m_{l,k} 
\right) 
(\cdots).
\label{ZZform2}
\end{equation}
From (\ref{weakZ2}) and (\ref{ZZform2}), 
the partition function $Z$ is obtained as
\begin{equation}
Z=\int_0^{2\pi}\frac{{\rm d}\theta}{2\pi} 
\exp\left( \sum_{l,k} b_{l,k} (\beta \Dbar)^{l+k}
{\rm e}^{{\rm i}(l-k)\theta} \right).
\label{weakZ}
\end{equation} 
For the dissipationless case ($K=0, \Dbar=\Delta$),
$b_{l,k}$ vanishes for $l+k\ge 2$ and the partition function 
is exactly calculated as
\begin{equation}
Z = \int_0^{2\pi} \frac{{\rm d}\theta}{2\pi} 
\exp\left(2\beta\Delta\cos\theta\right)=I_0(2\beta\Delta),
\end{equation}
where $I_0(z)$ is the Modified Bessel function. Of course,
this result is also obtained
by usual treatment for the tight-binding model of a free particle.

\begin{figure}[tb]
\hfil
\epsfile{file=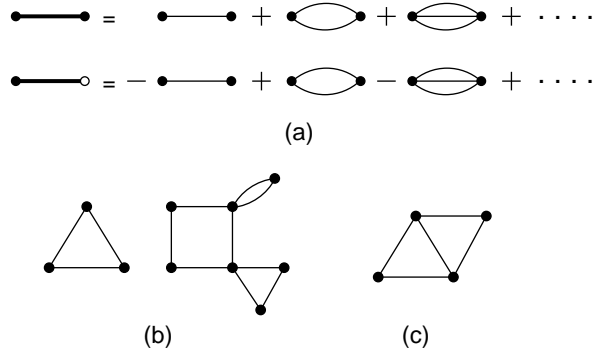,scale=0.5}
\hfil
\caption{The diagrammatic graphs representing the expansion
in (\ref{fij1}) and (\ref{fij2}) are drawn in (a). 
The graphs in both (b) and (c) 
are representative cluster integrals: (b) each charge 
has even bonds, while (c) includes charges with odd bonds.}
\label{graph2}
\end{figure}

So far, the obtained expression for $Z$ is exact. In the following
discussion, the so-called `ring approximation' is introduced.
First, we expand $f_{ij}^{\sigma_i\sigma_j}$ as
\begin{eqnarray}
f_{ij}^{++} = f_{ij}^{--} &=& \sum_{n=1}^{\infty}
\frac{\{\phi(\tau_i-\tau_j)\}^n}{n!}, 
\label{fij1} \\
f_{ij}^{+-} &=& \sum_{n=1}^{\infty}
\frac{\{-\phi(\tau_i-\tau_j)\}^n}{n!}.
\label{fij2}
\end{eqnarray}
Graphical representations of the expansion is given in 
Fig.~\ref{graph2}(a), where the bond denoted with $n$ thin lines 
denotes $\phi(\tau_i-\tau_j)^n$.
Here, we assume that (I) the dominant contribution 
comes from the clusters
in which each charge is combined by `even' bonds  
to other charges. Based on the assumption (I), we neglect other
clusters which do not satisfy this condition.
For example, we consider the graphs as shown in Fig.~\ref{graph2}(b),
and neglect the graphs as shown in Fig.~\ref{graph2}(c).
The assumption (I) is justified in the ring approximation
denoted later in detail. 
At this stage, however, in order to avoid confusion, we do not 
justify the assumption (I).
We mention this point after introducing the ring approximation.

The diagrams as shown in Fig.~\ref{graph2}(b)
have a crucial property that the cluster integral
on the graphs remains constant when the sign of each charge is reversed.
As a result, the cluster integral $l!k!b_{l,k}$ depends only on
the total charge number $l+k$. 
In this situation, $l!k!b_{l,k}$ can be denoted with $(l+k)!b_{l+k}$,
and the sum over $l$, $k$ under the constraint $l+k=m$ 
in the partition function (\ref{weakZ}) gives
\begin{equation}
Z = \int_0^{2\pi}\frac{{\rm d}\theta}{2\pi} 
\exp \left(\sum_{m=1}^{\infty} b_m \xi^m \right).
\label{onecomponent}
\end{equation}
Here $\xi=2\beta\Dbar\cos\theta$ is regarded as
a fugasity of `one-component' classical charges. 
Because the remaining calculation follows the usual procedure
of the classical imperfect gas as given in ref.~\citen{Mayler40},
we briefly note the results.
The diagrams, in which all charges are more than singly connected,
are called irreducible diagrams. The irreducible integrals
$\beta_l$ are defined on the irreducible diagrams with the size 
$(l+1)$ as
\begin{equation}
\beta_l = \frac{1}{l!} 
\sum_{{\rm irreducible} \atop{\rm clusters} } 
\langle \prod_{i,j}( \phi(\tau_i-\tau_j) )^{n_{i,j}} \rangle,
\end{equation}
where the product is taken over all bonds in a cluster, and
$n_{i,j}$ is the number of bonds combining $i$-th and $j$-th
charges. Every cluster integral $b_l$
can be expressed as a sum of terms, each of which is a numerical
coefficient multiplied by a product of powers of the reduced integrals
$\beta_l$ as
\begin{equation}
\begin{array}{ll}
b_1 = 1, & b_3 = \frac12 \beta_1^2 + \frac13 \beta_2, \\
b_2 = \frac12 \beta_1, & b_4 = \frac23 \beta_1^3 + \beta_1\beta_2
+ \frac14 \beta_3.
\end{array}
\label{betabrelation}
\end{equation}
In general, it is proved that the equation for $b_l$ is 
\begin{equation}
b_l = \frac{1}{l^2} \sum_{\{m_k\}'} \prod_k 
\frac{(l\beta_k)^{m_k}}{m_k!},
\label{clustbeta}
\end{equation}
where the prime in $\{m_k\}'$ denotes the summation subject to 
the constraint
\begin{equation}
\sum_{k=1}^{l-1} k m_k = l-1.
\end{equation} 

We also define a particle density $n$ by
\begin{equation}
n = \sum_{m=1}^{\infty} m b_m \xi^m.
\label{nexpand}
\end{equation}
This value corresponds to the average of the hopping number
in the imaginary-time path $q(\tau)$ in the original action, 
and qualitatively gives an effective hopping amplitude through 
$\Deff^{-1}\simeq \beta/n$.
By using $\beta_l$ and $n$, the partition function $Z$ can be
expressed in a closed form. To see this, we expand $\xi$ by $n$ as
\begin{equation}
\xi = a_1 n + a_2 n^2 + \cdots,
\label{xiexp}
\end{equation}
and determine $a_1$, $a_2$, $\cdots$ so that (\ref{nexpand})
is satisfied. Thus, we obtain
\begin{equation}
\begin{array}{ll}
a_1 = b_1^{-1} = 1,  &  a_3=8b_2^2-3b_3, \\
a_2 = -2b_2, & a_4 = -40b_2^3 + 30b_2b_3 - 4b_4.
\end{array}
\label{invcond}
\end{equation}
From (\ref{betabrelation}), 
these results can be rewritten in terms of the $\beta_l$'s as
\begin{equation}
\begin{array}{ll}
a_1 = 1, & a_3 = - (\beta_2 - \frac12\beta_1^2), \\
a_2 = -\beta_1, & a_4 = - (\beta_3 - \beta_1\beta_2+\frac16\beta_1^3).
\end{array}
\label{abetarelation}
\end{equation}
In general, $\xi$ is expressed by the particle density $n$ as
\begin{equation}
\xi = 2\beta\Dbar\cos\theta = n \exp\left( -
\sum_{l=1}^{\infty} \beta_l n^l \right).
\label{nxirelation}
\end{equation}
It can be easily checked that the first four terms in 
expansion of (\ref{nxirelation}) gives (\ref{abetarelation}).
The partition function (\ref{onecomponent}) 
is expressed in terms of $\beta_l$ and $n$ by utilizing (\ref{xiexp})
and (\ref{abetarelation}). As a result we obtain 
\begin{eqnarray}
& & Z = \int_0^{2\pi} \frac{{\rm d}\theta}{2\pi} \exp[U(n)], 
\label{ZZZ} \\
& & U(n) = n \left(1-\sum_{l=1}^{\infty} \frac{l}{l+1}\beta_l n^l
\right).
\label{Udef}
\end{eqnarray}
When we introduce 
\begin{equation}
Q = \sum_{l=1}^{\infty} \beta_l n^l,
\end{equation}
the relation (\ref{nxirelation}) and the exponent $U(n)$ is 
expressed in simple forms as
\begin{eqnarray}
& & \xi = n {\rm e}^{-Q(n)} \\
& & U = n + \int_0^n {\rm d}n \, Q - nQ.
\label{form2}
\end{eqnarray}

\begin{figure}[tb]
\hfil
\epsfile{file=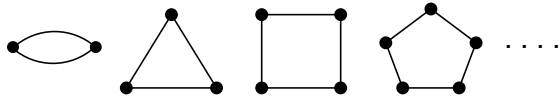,scale=0.6}
\hfil
\caption{The ring irreducible graphs considered in
the ring approximation.}
\label{graph3}
\end{figure}
At this stage, we introduce the approximation by assuming
that only the ring irreducible diagrams as shown in Fig.~\ref{graph3}
are dominant. This approximation is called the `ring approximation'.
Then, the irreducible integrals $\beta_l$ is expressed
by the Fourier component $\phi({\rm i}\omega_m)$ as
\begin{equation}
\beta_l = \frac{1}{2\beta^{l+1}} \sum_{{\rm i}\omega_m} 
\phi({\rm i}\omega_m)^{l+1},
\end{equation}
and thus we obtain
\begin{equation}
Q = \frac{1}{\beta^2} \sum_{\omega_m > 0}
\frac{n\phi({\rm i}\omega_m)^2}{1-n\phi({\rm i}\omega_m)/\beta}.
\label{form3}
\end{equation}
Here, the sum is taken over the Matsubara frequency, 
$\omega_m=2\pi m/\beta$. The justification of the ring approximation
is not clear. It, however, can be proved that this approximation
is equivalent to the Debye-H\"uckel theory, which has 
been considered to calculate thermodynamic quantities 
in strong electrolytes.~\cite{Landau80} In the Debye-H\"uckel theory,
the screening effects due to the other charges are described by
a screening potential $\vphi(\tau)$. (The explicit form
of the screening potential is given in (\ref{screenpotdef}).)
These screening effects are essential to study thermodynamic
quantities at low-temperatures in the original model.
In the ring approximation, the assumption (I)
adopted in the previous discussion is automatically satisfied,
because the neglected diagrams by the assumption (I) are 
also removed in the ring approximation.

We study the condition, in which the ring approximation is justified.
From (\ref{form3}), it is expected that the ring approximation is
certainly valid as long as the expansion quantity 
$n \phi(\omega_m)/\beta$ is small enough at 
$\omega_m\sim 1/\bar{\tau}$, where $\bar{\tau}
\sim \beta/n \sim \Deff^{-1}$ 
is a average distance between two particles. 
For the exponent $0<s<2$ in the spectral density, 
this condition corresponds to
\begin{equation}
\delta_s \left(\frac{\Deff}{\widetilde{\omega}}\right)^{s-1} 
\ll 1.
\end{equation}
In the ohmic damping case ($s=1$, $\delta_s = K$), 
this condition is nothing but the weak-damping condition
$K\ll 1$, where we expect that the
approximation gives reliable results in the weak-damping region.

Next, we calculate the optical conductivity.
The correlation function $\widetilde{\Lambda}_1(\tau)$ in
(\ref{lambda1}) is expressed by the cluster integrals as
\begin{eqnarray}
\widetilde{\Lambda}_1(\tau) &=& \frac{2\Dbar^2}{Z}
\int_0^{2\pi} \frac{{\rm d}\theta}{2\pi} \exp\left(
\sum_{l,k} b_{l,k} (\beta \Dbar)^{l+k} {\rm e}^{{\rm i}\theta
(l-k)} \right) \nonumber \\
&\times & \sum_{l,k} B_{l,k} (\beta\Dbar)^{l+k} 
{\rm e}^{{\rm i}\theta (l-k)}.
\label{onaka}
\end{eqnarray}
Here, $B_{l,k}$ is the cluster integral of the diagram
including two `fixed' charges at $\tau'=0,\tau$,
and $l$ and $k$ are the numbers of additional
positive and negative charges. 
Note that diagrams in which the two `fixed' charges are 
separated into two clusters do not contribute to the
optical conductivity, because such terms are independent of $\tau$.

\begin{figure}[tb]
\hfil
\epsfile{file=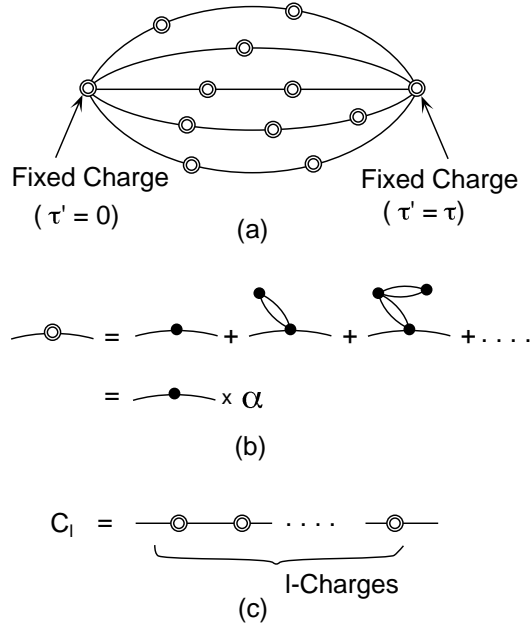,scale=0.55}
\hfil
\caption{(a) The representative diagram relevant to the optical 
conductivity in the ring approximation. The renormalization
of the fugasity $\xi$ is denoted with the sum of diagram (b).
The diagram (a) can be divided into the lines with $l$ 
charges as shown in Fig.~(c). }
\label{graph4}
\end{figure}
In the ring approximation, (\ref{onaka}) is reduced to the
partition function of one-component charged particles. 
Thus, we obtain
\begin{equation}
\widetilde{\Lambda}_1(\tau) = \frac{1}{2\beta^2 Z} \int_0^{2\pi}
\frac{{\rm d}\theta}{2\pi} 
\frac{\xi^2}{\cos^2\theta} 
\exp \left( \sum_{m=1}^{\infty}
b_m \xi^m \right) \sum_{m=1}^{\infty} B_m \xi^m,
\label{Lambda1cal}
\end{equation}
where $\xi = 2\beta\Dbar\cos\theta$.
Since the sum of $b_m \xi^m$ gives $U$ as shown in (\ref{onecomponent}) 
and (\ref{ZZZ})-(\ref{Udef}), 
the remaining problem is the sum of $B_m \xi^m$,
which comes from the cluster integrals of the diagrams including two
fixed charges. In the ring approximation, only
the diagrams shown in Fig.~\ref{graph4}~(a) are relevant. 
The double circle in the diagram denotes
the sum of the diagrams as shown in Fig.~\ref{graph4}~(b), and
renormalize the fugasity $\xi$ as 
\begin{equation}
\xi \rightarrow \xi \sum_{m=0}^{\infty} m b_m \xi^{m-1} 
= \xi \frac{{\rm d}U}{{\rm d}\xi}.
\end{equation}
By using (\ref{Udef}), it is proved that
the renormalization is given by $\xi\rightarrow n$, 
where $n$ is the particle density. 

Finally, we consider
the sum of the diagrams represented by Fig.~\ref{graph4}~(a).
The cluster integrals including two `fixed' charges
are summed up as
\begin{eqnarray}
\sum_{m=0}^{\infty} B_m n^m &=& \sum_{m=0}^{\infty} 
\frac{n^m}{m!} \sum_{\{m_l\}'} 
\frac{m!}{\dis{\prod_{l=0}^{m}m_l!(l!)^{m_l}}} 
\prod_{l=0}^{m} (C_l l!)^{m_l} \times (-1)^{m_l} m_l! 
\label{cluster0}
\\
&=& \exp\left(-\sum_{l=0}^{\infty} C_l n^l \right).
\label{cluster1}
\end{eqnarray}
Here, $C_l$ denotes the sum shown in Fig.~\ref{graph4}(c),
and is obtained as
\begin{equation}
C_l = \frac{1}{l!} \times \frac{l!}{\beta^{l+1}} \sum_{{\rm i}\omega_m}
\phi({\rm i}\omega_m)^{l+1} {\rm e}^{-{\rm i}\omega_m \tau}.
\label{cluster2}
\end{equation}
From (\ref{Lambda1cal}) and (\ref{cluster1})-(\ref{cluster2}), 
the equation ({\ref{weaklambda1}) is derived. 
Similarly, $\widetilde{\Lambda}_2(\tau)$ is derived by
removing the last term $(-1)^{m_l}$ in (\ref{cluster0}), and by
replacing the electroneutrality condition (\ref{constraint}) 
with (\ref{constraint2}).

\section{Low-Temperature Expansions}\label{app3}
In this appendix, we discuss the low-temperature
expansions for the specific heat and the DC conductivity 
in the case of the ohmic damping.
This expansion is obtained by asymptotic expansions 
in terms of the inverse temperature $\beta=1/T$.
We begin with the partition function $Z$ expressed by
\begin{equation}
Z =\int_{-\pi}^{\pi} \frac{{\rm d}\theta}{2\pi}
{\rm e}^{U(\theta)}.
\label{appint}
\end{equation} 
Here, $U(\theta)=U(n(\theta))$ is given by
\begin{equation}
U(n) = n + \log \Gamma (Kn+1)-Kn\psi(Kn+1),
\end{equation}
where $\Gamma(z)$ is the gamma function, and 
$\psi(z)$ is the polygamma function.
At low temperatures, the integral in (\ref{appint}) is
determined by the contribution around $\theta=0$.
We expand $U(\theta)$ as
\begin{equation}
U(\theta) \simeq U(0) + \frac12 U''(0) 
\theta^2 + {\cal O}(\theta^4) , \label{Uexpansion}
\end{equation}
and the integral (\ref{appint}) is replaced with 
the Gaussian integral by (\ref{Uexpansion}), approximately.
By using $U''(0)=-n_0$ and asymptotic forms of $\Gamma(z)$ and
$\psi(z)$, we obtain
\begin{equation}
Z = (1-K)n_0 + {\rm const.} + \frac{1}{6Kn_0} + {\cal O}(n_0^{-2}),
\label{Zasympt}
\end{equation}
where $n_0=n(\theta=0)$. We should note that the asymmetry
part $U''''(0)$ also gives a term of order of $1/n_0$.
However, the term takes a finite value for $K\rightarrow 0$,
and for weak damping $(K\ll1$), and this contribution is
less than the third term of (\ref{Zasympt}).
From (\ref{Qnexp}) and (\ref{nthetaexp}), $n_0$ is determined by
\begin{eqnarray}
& & 2\beta\Dbar = n_0 {\rm e}^{-Q(n_0)} 
\label{gakkai0} \\
& & Q(n_0)=K(\psi(Kn_0+1)+\gbar).
\label{gakkai-1}
\end{eqnarray}

Next, we calculate the asymptotic form of (\ref{gakkai0}).
By using $\Dbar=\Delta(\beta \omega_c/2\pi)^{-K}$ and
$\Deff = \Delta(\Delta/\omega_c)^{K/(1-K)}$,
the lefthand side of (\ref{gakkai0}) is represented as 
$4\pi(\beta\Deff/2\pi)^{1-K}$. For $\beta\rightarrow\infty$
($n\rightarrow\infty$), the righthand side of (\ref{gakkai0})
is expanded as
\begin{equation}
n_0 {\rm e}^{-Q(n_0)} = (K{\rm e}^{\gbar})^{-K} n_0^{1-K} 
\left( 1 - \frac{1}{2n_0}+ \frac{1}{12Kn_0^2} + {\cal O}(n_0^{-3})
\right).
\label{gakkai1}
\end{equation}
We substitute $n_0=a_1 \beta + a_0 + a_{-1}/\beta+\cdots$ to
(\ref{gakkai1}), and determine the constants $a_1$, $a_0$,
$a_{-1}$ so that (\ref{gakkai1}) gives the lefthand side
of (\ref{gakkai0}). Thus, we obtain
\begin{equation}
n_0=2\beta c\Deff +\frac{1}{2(1-K)} - \frac{1}{24K\beta c\Deff}
+ {\cal O}(\beta^{-2}),
\label{fififi}
\end{equation}
where $c=(4\pi K{\rm e}^{\gbar})^{K/(1-K)}$. From (\ref{Zasympt}) 
and (\ref{fififi}),
we obtain the low-temperature expansion 
of the partition function, (\ref{Zasymp}).

In order to calculate
$\sigma_{\rm DC}$, it is sufficient to consider the long-time
behavior of $S(t;\theta)$ and $R(t;\theta)$ defined 
in (\ref{Sweak})-(\ref{Rweak}). From (\ref{Rweak}) and (\ref{Stinf2}),
we obtain for $t\rightarrow \infty$
\begin{eqnarray}
& & S(t;\theta) = \frac{2\pi K T}{\gamma(\theta)} = 
\frac{1}{n(\theta)},
\label{Stinf} \\
& & R(t;\theta) = \pi K {\rm e}^{-\gamma(\theta)t}.
\label{Rtinf}
\end{eqnarray}
Here, we have used $\gamma(\theta)=2\pi Kn(\theta)/\beta$.
From (\ref{weakformalism}), the real-time correlation function 
in the long-time limit is obtained as
\begin{equation}
\Lambda(t) = \frac{e^2 a^2}{Z} \int_{-\pi}^{\pi}
\frac{{\rm d}\theta}{2\pi} \frac{n^2}{\beta^2\cos^2 \theta}
{\rm e}^{U(\theta)} \left[ {\rm e}^{-1/n} + 
{\rm e}^{1/n} \cos 2\theta \right]
\pi K {\rm e}^{-\gamma(\theta)t}.
\end{equation}
Then, the DC conductivity is calculated as
\begin{eqnarray}
\sigma_{\rm DC} &=& \lim_{\omega\rightarrow 0}
\frac{1}{\omega} {\rm Im} \int_0^{\infty} {\rm d}t \,
{\rm e}^{{\rm i}\omega t} \Lambda(t)  \nonumber \\
&=& \frac{\sigma_{\rm DC}^0}{2Z} \int_{-\pi}^{\pi} \frac{{\rm d}\theta}{2\pi}
{\rm e}^{U(\theta)} \left[
\frac{{\rm e}^{-1/n}-{\rm e}^{1/n}}{\cos^2 \theta} 
+ 2 {\rm e}^{1/n} \right],
\label{last}
\end{eqnarray}
where $\sigma_{\rm DC}^0 = e^2a^2/2\pi K$. The low-temperature 
expansion is obtained by expanding (\ref{last}) over
$1/n(\ll 1)$ as
\begin{equation}
\sigma_{\rm DC} = \frac{\sigma_{\rm DC}^0}{Z} \int_{-\pi}^{\pi}
\frac{{\rm d}\theta}{2\pi} \left[ 1 + \frac{1}{2n^2} -
\frac{1}{n}\theta^2 + {\cal O}(n^{-2},\theta^4) \right] 
{\rm e}^{U(0)+U''(0)\theta^2/2}.
\end{equation}
By using the Gaussian integral formulas and an asymptotic form
of the partition function
\begin{equation}
Z = \int_{-\infty}^{\infty} \frac{{\rm d}\theta}{2\pi} 
{\rm e}^{U(0)+U''(0)\theta^2/2},
\end{equation}
we obtain
\begin{equation}
\sigma(\omega) = \sigma_{\rm DC}^0 \left( 1-\frac{1}{2n^2}
+ {\cal O}(n^4) \right).
\label{last2}
\end{equation}
From (\ref{fififi}) and (\ref{last2}), we obtain
(\ref{dccondT}).

%

\begin{thebibliography}{99}
\bibitem{footnote1} In the low-temperature region,
this simplification may be problematic in several systems.
For details, see \S~\ref{OhmicGreen}.
\end{thebibliography}
\end{document}